\documentclass[aip,reprint,letterpaper,numerical]{revtex4-1}

\usepackage{color} 
\usepackage{geometry}
 \usepackage{graphicx} 
 \usepackage{amssymb}
\usepackage{bm} 
\usepackage{amsmath} 
\newcommand{\mat}[1]{\mathrm{\underline{\underline{#1}}}}

\begin{document} 

\title{Interplay between hydrodynamics and the free energy surface in
the assembly of nanoscale hydrophobes}

\author{Joseph A. Morrone} 
\affiliation{Department of Chemistry, Columbia University, 3000 Broadway, New York, NY 10027}

\author{Jingyuan Li} 
\altaffiliation{Present address: Chinese Academy of Sciences Key Lab for Biomedical Effects of Nanomaterials and Nanosafety, National Center for Nanoscience and Technology, Institute of High Energy Physics, Chinese Academy of Sciences, Beijing 100190, China}
\affiliation{Department of Chemistry, Columbia  University, 3000 Broadway, New York, NY 10027}

\author{B. J. Berne}
\email{bb8@columbia.edu} 
\affiliation{Department of Chemistry, Columbia University, 3000 Broadway, New York, NY 10027}

\begin{abstract} Solvent plays an important role in the
    relative motion of nanoscopic bodies, and the study of such
    phenomena can help elucidate the mechanism of hydrophobic
    assembly, as well as the influence of solvent-mediated effects on
    \emph{in vivo} motion in crowded cellular environments.  Here we
    study important aspects of this problem within the framework of
    Brownian dynamics.  We compute the free energy surface that the
    Brownian particles experience and their hydrodynamic interactions
    from molecular dynamics simulations in explicit solvent. We find
    that molecular scale effects dominate at short distances, thus
    giving rise to deviations from the predictions of continuum
    hydrodynamic theory. Drying phenomena, solvent layering, and
    fluctuations engender distinct signatures of the molecular
    scale. The rate of assembly in the diffusion-controlled limit is
    found to decrease from molecular scale hydrodynamic interactions,
    in opposition to the free energy driving force for hydrophobic
    assembly, and act to reinforce the influence of the free energy
    surface on the association of more hydrophilic bodies.
\end{abstract}

\maketitle

\section{Introduction}

The hydrophobic effect plays a fundamental role in biology and
nanoscale engineering.  Great strides have been made in our
understanding of the mechanism of hydrophobic assembly by elucidating
the behavior of solvent near interfaces of variable length
scale.~\cite{pangali79,Chandler:2005p75,Berne:2009dg,Jamadagni:2010bc,WALLQVIST:1995uw,hummer96,Lum:1999p624,huang02,TenWolde:2002p71,Margulis:2003wo,ruhong04,Huang:2005ho,nicolas07,mittal2008,Willard:2008p84,Sarupria:2009dq,godawat09,patel10}
In addition, the hydrophobic effect is crucial to understanding the
nature of surface friction in nanoscale fluid
transport.~\cite{Sendner:2008tg,Sendner:2009gi,Thomas:2009tj,Kalra:2010p231,Falk:2010cr,Bocquet:cy}
Surface friction characterizes the boundary condition in the continuum
description for fluid flow across interfaces of varying
hydrophobicity.

Despite recent efforts to improve our understanding of the hydrophobic
effect, to our knowledge the role of hydrodynamic interactions in
hydrophobic assembly has not been considered.  Within the framework of
Brownian dynamics, hydrodynamic interactions are incorporated in the
frictional force by means of a resistance (friction) tensor that is
dependent upon the degrees of freedom of the Brownian
bodies.~\cite{mccammon78} Hydrodynamic interactions are typically
treated in the simulation of colloidal suspensions within the
continuum description of low Reynolds number
flow.~\cite{Brady:1988tz,Brady:1988p970,chatterji05,padding06}
Recently, polydisperse colloidal simulations intended to mimic
cytoplasm reported that hydrodynamic interactions give rise to the
slow diffusion of macromolecules observed in crowded cellular
environments.~\cite{Ando:2010p99}

For small separations, the continuum description is dominated by
lubrication effects, which yields a friction constant that diverges as
two bodies come into contact.  Such frictional forces arise from the
assumption that there is a fluid element present even as the spacing
approaches zero.  Naturally, this description breaks down when
molecular scale effects become important at small length scales.  The
continuum description of the Navier-Stokes equation is expected be
valid down to length scales of approximately $1 - 2$
nm.~\cite{Thomas:2009tj,Bocquet:cy}  Therefore, molecular-scale
effects should be considered when only a few solvent layers separate
the solute.

The spatially dependent friction tensor can be extracted from explicit
molecular dynamics simulations where all bath degrees of freedom are
included.  This may be achieved in either the Brownian limit or when
memory effects are
included.~\cite{Berne:1970p90,Straub:1987p76,Straub:1990p65,Espanol:1993p63,Bocquet:1994p317,Bocquet:1997p103}
To our knowledge, the friction tensor between two bodies has only been
previously computed in Lennard-Jones
fluids.~\cite{Straub:1990p65,Bocquet:1997p103,Lee:2010p307} Presently
we consider the spatial dependence of the friction constant as
function of the distance between two non-polar spherical bodies in
explicit water.  The spatial friction, in concert with the
(non-hydrodynamic) potential of mean force along the relative
direction, facilitates the modeling of two-body assembly in the
Brownian limit and an assessment of the importance of hydrodynamic and
non-hydrodynamic forces.

This work aims to elucidate the role that hydrodynamics plays in
hydrophobic assembly.  We take as our model solute two fullerenes,
either two C60s or two C240s. Unsubstituted fullerenes are insoluble
in water,~\cite{fulsol} and thus, to our knowledge there are no
experimental data on their transport in water. Nevertheless they are a
useful model for theoretically probing the transport of nanoscale
bodies.  The hydrophobic collapse of two bodies in explicit
water~\cite{Margulis:2003wo,Huang:2005ho} and pathways for
self-assembly of hydrophobic spheres in a coarse grained solvent
model~\cite{Willard:2008p84,TenWolde:2002p71} have been previously
studied.  The potential of mean force of two C60 molecules in water
has also been computed.~\cite{Hotta:2005bp,Makowski:2009p61} However,
the interplay between the free energy surface and hydrodynamic
interactions has not yet been considered.  As we will show, solvent
fluctuations and drying phenomena manifest themselves as hydrodynamic
interactions within the Brownian framework.  We find that because slow
relaxation times are associated with the solvent at the drying
transition the separation of solute and solvent time scales required
by a Brownian description of hydrophobic assembly may not always be
satisfied.

The solute-solvent interactions are described by both attractive and
purely repulsive potentials.  For the purely repulsive model of the
fullerene we observe a drying transition,~\cite{WALLQVIST:1995uw} one
which is more pronounced for C240 than C60, a finding which is in
agreement with the known length-scale dependence of drying
phenomena.~\cite{Lum:1999p624,Margulis:2003wo} The attractive spheres
do not dry as they approach each other and water is expelled from the
inter-solute region by means of steric repulsion.  We find that the
spatial dependence of the friction constant is quite different in each
of these cases.  In particular, drying phenomena and solvent layering
engendered by attractive potentials are shown to possess distinct
signatures in their respective friction tensor.  Hydrodynamic
interactions at the molecular scale are related to the variability of
solvent density, fluctuations, and relaxation times in the
inter-solute region as two bodies approach.

The rate constant for assembly is computed via a Smoluchowski
Analysis.~\cite{emeis70,deutch73,wolynes76,Northrup:1979wr,calef83,Straub:1990p65}
We find that for two body assembly, the inclusion of hydrodynamic
interactions decreases the rate of reaction in the diffusion
controlled limit by approximately $30 - 40\%$.  In the case of ideal
hydrophobic assembly (that is when no solute-solvent attraction is
present) the frictional and mean force can be said to have opposite
effects.  When solute-solvent attraction is included, they tend to
reinforce each ether's impact on the rate of assembly.

This paper is organized as follows.  Sec. \ref{sec:method} reviews the
formalism of hydrodynamic interactions and Brownian motion. Simulation
details are given in Sec. \ref{sec:details}.  Results are given for the
friction coefficient on a single body in Sec. \ref{sec:single} and are
compared with the Stokes-Einstein relation with periodic boundary
corrections. In Sec. \ref{sec:two} an analysis of the assembly of two
non-polar solutes within the framework of Brownian motion is
presented.  Discussions and conclusions are given in Sec. \ref{sec:conc}.

\section{Langevin dynamics and hydrodynamic
  interactions} \label{sec:method} In this work we seek to
characterize solvent-mediated effects on two spherical bodies.  The
Brownian limit is considered such that the bodies evolve on a much
slower time scale than the solvent bath.  The dynamics of such a
system can be described by an equation of motion that includes
contributions from three forces, a frictional force, a mean force,
and an uncorrelated, Gaussian random force.  Due to the symmetry of
the problem, the relative distance $r=\left| \vec{r}_2 - \vec{r}_1
\right|$ between the spheres is the only coordinate necessary to
characterize the approach of the two solute bodies.  In the present
study we concentrate on this direction, and will consider the friction
along other degrees of freedom in future work.  When hydrodynamic
interactions are included in the frictional force, the Langevin
equation for the motion of the relative coordinate is given
by,~\cite{mccammon78}
\begin{align} \mu \ddot{r} &= -\nabla \mathcal{W}(r) - \zeta(r)\dot{r}
+ R(r,t)\label{eqn:rel_lange} \; ,
\end{align} where $\mu$ is the reduced mass and $\mathcal{W}(r)$ is
the potential of mean force.  In the Brownian limit the solvent
(hydrodynamic) time scale is separable from the motion of the solute
and so the friction tensor may be computed at fixed solute
positions.~\cite{dhont} This is analogous to the Born-Oppenheimer
approximation, which assumes a similar separation of time scales
between nuclear and electronic degrees of freedom.  The
(non-hydrodynamic) free energy profile contains contributions from
both direct solute-solute interactions and induced solvent-mediated
forces.  In the limit of large friction the term on the left-hand side
of Eqn. \ref{eqn:rel_lange} may be neglected.  The friction coefficient
$\zeta(r)$ can be determined from the full two-body tensor as reviewed
in the Appendix.  The random force, $R(r,t)$, is Gaussian,
uncorrelated noise with zero mean and a variance that is related to
the friction coefficient via the fluctuation-dissipation theorem,
\begin{align} \left< R(r,t) R(r,0) \right> &= 2 k_b T \zeta(r)
\delta(t) \; .
\end{align} As indicated by Eqn. \ref{eqn:rel_lange}, when hydrodynamic
interactions are included the friction coefficient depends on the
distance between spheres.  If hydrodynamic interactions are ignored
then $\zeta(r) \rightarrow \zeta_0$, where $\zeta_0$ is taken to be
the friction coefficient on the relative coordinate at infinite
separation where no spatial dependence is assumed.

Hydrodynamic interactions that yield the spatial dependence of the
friction coefficient are typically computed by means of approximate
solutions of the linearized Navier-Stokes equation valid for
incompressible flow at low Reynolds number.  A ubiquitous formulation
of hydrodynamics frequently utilized in colloidal simulations is
Stokesian dynamics.~\cite{Brady:1988tz,Brady:1988p970}  Stokesian
dynamics interpolates between an exact two-body solution for short
range interactions~\cite{Jeffrey:1984tu} and the long range many-body
result derived from the Rotne-Prager tensor as applied to systems with
periodic boundary conditions.~\cite{Beenakker:1986tz}.  Other
techniques may also be utilized to compute continuum hydrodynamic
interactions~\cite{Chen:1998vv,chatterji05,padding06,Nakayama:2008p587}
and the continuum formulation has been recently employed as a
framework for coarse-grained solvent models.~\cite{chu09} In this work
we will compare our molecular scale results to both the expression of
Ref. \onlinecite{Jeffrey:1984tu} and the Oseen tensor computed with
periodic boundary conditions.~\cite{Hasimoto:1959p589,Lindbo:2010ha}
The Oseen tensor is the leading order term of the Rotne-Prager
expression~\cite{dhont}.

The continuum approach breaks down for small particle separations, and
in this work we seek to determine hydrodynamic interactions at the
molecular scale.  The solvent bath is treated by explicit Newtonian
(microcanonical) molecular dynamics simulation. All bath degrees of
freedom and microscopic details are present in our simulation.  The
friction coefficient is related to the microscopic fluctuations in the
linear response regime by means of a Green-Kubo relation.  This
equates the friction tensor to the correlation function of the
fluctuations of the total force on each Brownian body, $\delta F_i =
F_i - \left<F_i\right>$.~\cite{bernebook}.
\begin{align} \mat{\zeta}_{ij} = \frac{1}{k_bT} \int\limits_0^\infty
\mathrm{d}t \; \lim\limits_{N \rightarrow \infty} \left< \delta
\vec{F}_i (t) \delta \vec{F}_j(0) \right> \label{eqn:greenkubo}
\end{align} Where $\zeta_{ij}$ is a component of the friction tensor
(see Appendix).  Unfortunately, this formula may only be directly
applied in both the Brownian limit ($M_\mathrm{solute} \rightarrow
\infty$) and in the thermodynamic limit where the number of solvent
molecules, $N$, approaches
infinity.~\cite{bernebook,Bocquet:1994p317,Bocquet:1997p103}.
Although the former requirement may be satisfied by fixing the
positions of the Brownian particles, the latter requirement gives rise
to subtle complications in a finite simulation.  Utilizing the
techniques developed by Bocquet \emph{et Ab.} Eqn. \ref{eqn:greenkubo} may
be employed to \emph{indirectly} compute the two-body friction
coefficient, that is the MD estimate of $\mat{\zeta}_{ij}$ can be
related to linear combinations of the submatricies of the friction
tensor.~\cite{Bocquet:1997p103} Similar complications are present in
the computation of the friction on a single
body.~\cite{Bocquet:1994p317} Further details are provided in the
Appendix.

\section{System setup and simulation details} \label{sec:details} 
\begin{figure*}
\begin{center}
\includegraphics[scale=0.75]{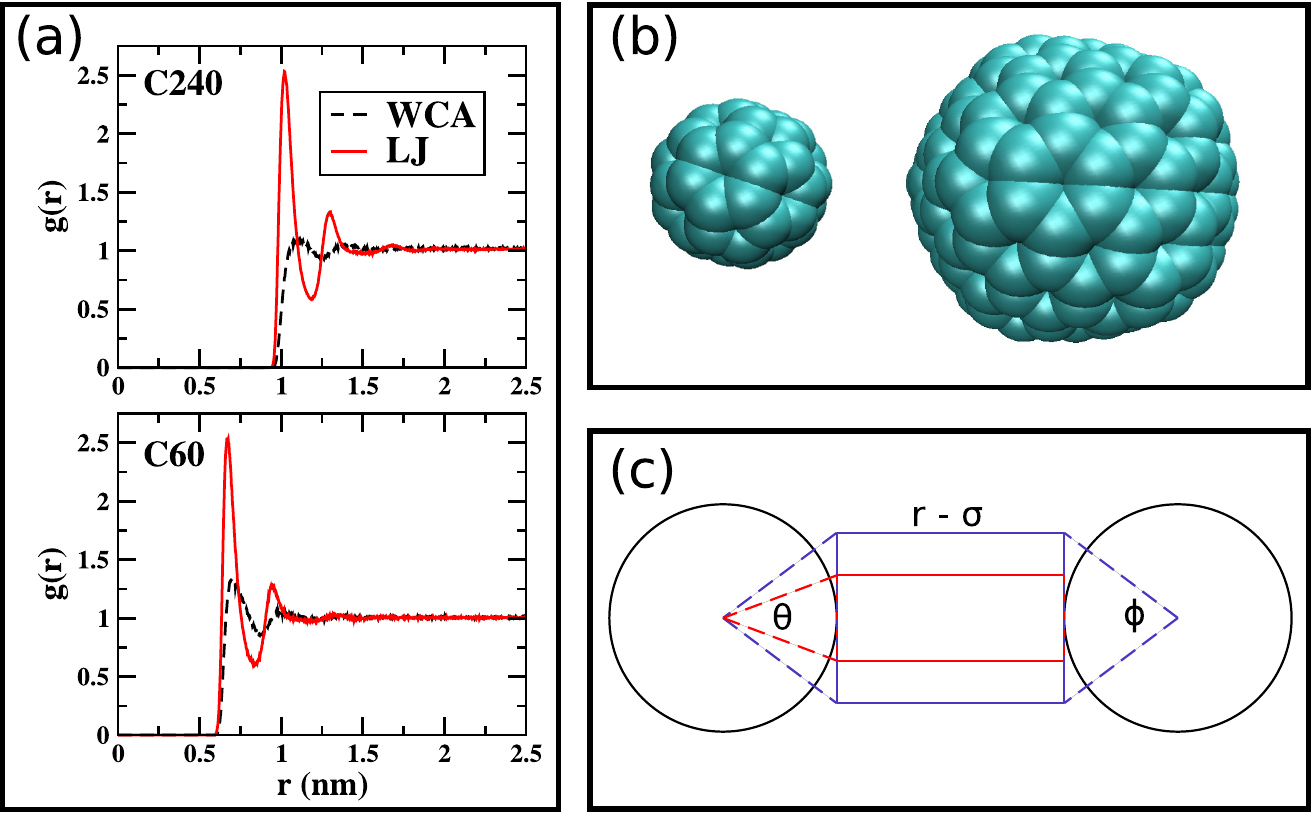}
\caption{(a) The pair-correlation function from the center of C240
(top panel) or C60 (bottom panel) to the oxygen site on the water
molecule is depicted for the cases when the solute-solvent potential
is purely repulsive (dashed black curve) or includes attraction (solid
red curve). (b) The two fullerenes utilized to model non-polar
spherical bodies, C60 (left) and C240 (right) are visualized with
VMD.~\cite{vmd96} The estimated diameter for each species is 1.0 nm
and 1.7 nm, respectively.  (c) Two spherical bodies of diameter
$\sigma$ are placed a distance $r$ from each center.  A cylindrical
probe volume (blue lines) is chosen such that its length is $r-\sigma$
and its diameter determined by the angle $\phi=90^\circ$.  This volume
can be further partitioned into an inner-tube (red lines) whose
diameter is determined by the angle $\theta=60^\circ$ and an outer
shell that is the difference between the two probe
volumes.} \label{fig:mixed}
\end{center}
\end{figure*}

In
order to gauge the impact of size effects, we utilize two fullerenes,
C60 and C240, as our model non-polar bodies.  The two molecules and
their relative sizes are depicted in panel (b) of Fig. \ref{fig:mixed}.
The solute-solvent interactions are described by two forms, a
Lennard-Jones (LJ) interaction with parameters $\sigma_\text{SS}=0.332
\text{ nm}$ and $\epsilon_\text{SS}=0.423 \text{ kJ/mol}$, and a
Weeks-Chandler-Andersen (WCA) truncation of this
potential~\cite{Weeks:1971us}.  As will be shown below, there is
vastly different behavior when the potential is attractive or purely
repulsive.  For the purpose of the friction analysis the fullerenes
are treated as spheres and the friction coefficient is computed on their
respective centers of mass. Based on the choice of repulsive core, the
estimated Van der Waals diameters for C60 and C240 are $\sigma=1.0$ nm
and $\sigma=1.7$ nm, respectively.

The friction is sensitive to periodic boundary conditions (see
Sec. \ref{sec:single}), and we utilize as large a system as is
computationally feasible.  For runs containing C60 the base box length
prior to equilibration is 5.0 nm and for C240 it is 6.0 nm.  In
systems containing two fullerenes, a set of calculations was run with
the bodies frozen and set at a fixed distance apart from each other.
The box length is extended by the separation between solute centers of
mass in the axial direction of the starting configuration of each
computation.  The number of water molecules utilized to represent the
solvent varies between approximately $5000 - 11000$ depending on the
size of the fullerene and the size of the separation.  Water is
modeled with the TIP4P potential~\cite{tip4p}. Snapshots~\cite{vmd96}
of two C240 molecules fixed at three different separations are shown
in Fig. \ref{fig:panels}.  Each system was equilibrated under NPT
conditions for 2 ns and NVT conditions for 1 ns. The barostat of
Berendsen~\cite{berendsen} and a stochastic velocity rescaling
thermostat~\cite{Bussi:2007p114} were employed during equilibration
runs to maintain a temperature of 300 K and a pressure of 1 bar.  In
order to compute the friction coefficient, data was collected from $10
- 18$ NVE runs of 4 ns in length at each fixed distance.  All
calculations were performed using GROMACS version
4.5.3.~\cite{gromacs4} As the friction is a probe of the solvent
momentum relaxation in the presence of the solute bodies, its
computation requires strict energy conservation in the microcanonical
ensemble. NVE runs were performed in double precision.
 
The computation of the spatially dependent friction coefficient
requires constant energy simulations and three dimensional periodic
boundary conditions are necessary for comparison with long range
periodic continuum hydrodynamics. Therefore, behavior at the drying
transition must be studied at constant volume and in the absence of an
interface.  The number of solvent molecules we choose is large enough
to accommodate the study of a dewetting transition and although the
behavior is modestly perturbed by this choice of ensemble, the
qualitative features of the friction at the critical distance for
dewetting are not expected to suffer.  As the two solutes are fixed
throughout all computations, direct solute-solute interactions are not
a component of the simulations described above.  Direct interactions
between fullerenes are included in the computation of the rate
constant in Sec. \ref{sec:rate} and the solute-solute potential has the
parameters $\sigma=0.35 \; \text{nm}$ and $\epsilon=0.276
\;\text{kJ/mol}$.

\section{Friction on a fullerene in solvent} \label{sec:single} The
strength of non-polar attractions play an important role in the degree
of particle hydrophobicity (see e.g. Ref. \onlinecite{huang02}).  This is
readily seen in panel (a) of Fig. \ref{fig:mixed}, where the pair
correlation function of the solvent with respect to the center of mass
of the fullerene is plotted.  Whereas the attractive fullerenes
exhibit significant structuring of the first solvation shell, the
water density is depleted near the surface of the repulsive
bodies. The pair correlation function for C60 is in good agreement
with previous studies of hydration around fullerenes.~\cite{weiss08}
Furthermore, depletion becomes more prominent with increasing solute
size, a finding that is in agreement with prior theoretical
predictions and simulation.~\cite{Lum:1999p624,mittal2008}

The friction coefficient of a single C60 and C240 molecule is computed
utilizing both the LJ and WCA solute-solvent potentials.  These values
may be related to the Stokes radius via the Stokes-Einstein relation
corrected for periodic boundary conditions
(PBC).~\cite{Hasimoto:1959p589,Yeh:2004p287}   As noted in the cited
works, PBC have a large impact on the measured friction.  To leading
order this correction, as formulated in Ref. \onlinecite{Yeh:2004p287},
is given by the following expression,
\begin{align} \frac{1}{\zeta_\text{iso}} = \frac{1}{\zeta_\text{PBC}}
+ \frac{\xi}{6\pi\eta L} \;\; ,
\end{align} where $\xi=2.837297$, $L$ is the periodic box length, and
$\eta$ is the shear viscosity.  The value of the viscosity of the
TIP4P model used in this work has been taken from the
literature.~\cite{viscoref}

Once the friction coefficient of the isolated sphere,
$\zeta_\text{iso}$, has been computed, it can be related to the Stokes
radius, which is given by,
\begin{align} r_\text{stokes} &= \frac{\zeta_\text{iso}}{c \pi \eta}
\; \; , \label{eq:stokes}
\end{align} where $c$ is equal to $4$ or $6$ for slip or stick
boundaries, respectively.  Studies relating Eqn. \ref{eq:stokes} to the
results of molecular dynamics simulation have appeared in the
literature~\cite{schmidt03,schmidt04,li09} and subtleties associated
with the appropriate hydrodynamic radius and boundary choice should be
addressed for a complete understanding of how this continuum
expression applies at the molecular scale. Furthermore, we note that
if our model solutes were perfect spheres (which they are not), then
the slip boundary condition holds unless solute-solvent attractions
are so large that solvent molecules become entrenched on the
surface.~\cite{schmidt04} Presently, we simply assume either a stick
or slip boundary condition and compute the stokes radii of the
fullerenes with the LJ and WCA solute-solvent potentials.  The results
are presented in Table \ref{tab:stokesrad} and compared with the estimated
van der Waals radii.  One can see that the stokes radii of the purely
repulsive spheres with slip boundary conditions are in very good
correspondence with the van der Waals radii.  As the molecules become
more attractive, the friction coefficient increases as does in this
sense the ``stickiness'' of the surface.  This is in agreement with
studies of water at interfaces, which have shown that there is a
correlation between hydrophobicity and the friction tangential to the
surface.~\cite{Sendner:2008tg,Sendner:2009gi} Furthermore, studies of
particle diffusion in a Lennard-Jones fluid have exhibited similar
trends as a function of solute-solvent
attraction.~\cite{schmidt04,li09}

In the following analysis, we determine the friction coefficient along
the relative coordinate in units of $\zeta_0 = \zeta_\text{PBC}/2$ for
each respective fullerene and solute-solvent potential. The two body
simulations utilize a periodic box of dimensions $(L+d) \times L
\times L$ and therefore this choice accounts for the impact of
periodic conditions.  Changes relative to $\zeta_0$ are a result of
hydrodynamic interactions and not PBC artifacts.

\begin{table*}
\begin{center}
\begin{tabular}{||c|c|c|c|c||} \hline & & Stokes radius (slip) &
Stokes radius (stick) & Estimated vdW radius \\ \hline C60 & WCA &
0.53 nm & 0.35 nm & 0.50 nm \\ C60 & LJ & 0.66 nm & 0.44 nm & 0.50 nm
\\ C240 & WCA & 0.84 nm & 0.56 nm & 0.85 nm \\ C240 & LJ & 1.06 nm &
0.70 nm & 0.85 nm \\ \hline
\end{tabular}
\caption{Table of computed Stokes radii for C60 and C240 with
differing solute-solvent interactions. } \label{tab:stokesrad}
\end{center}
\end{table*}

\section{Two-body relative friction} \label{sec:two} 
\begin{figure*}
\begin{center}
\includegraphics[scale=0.33]{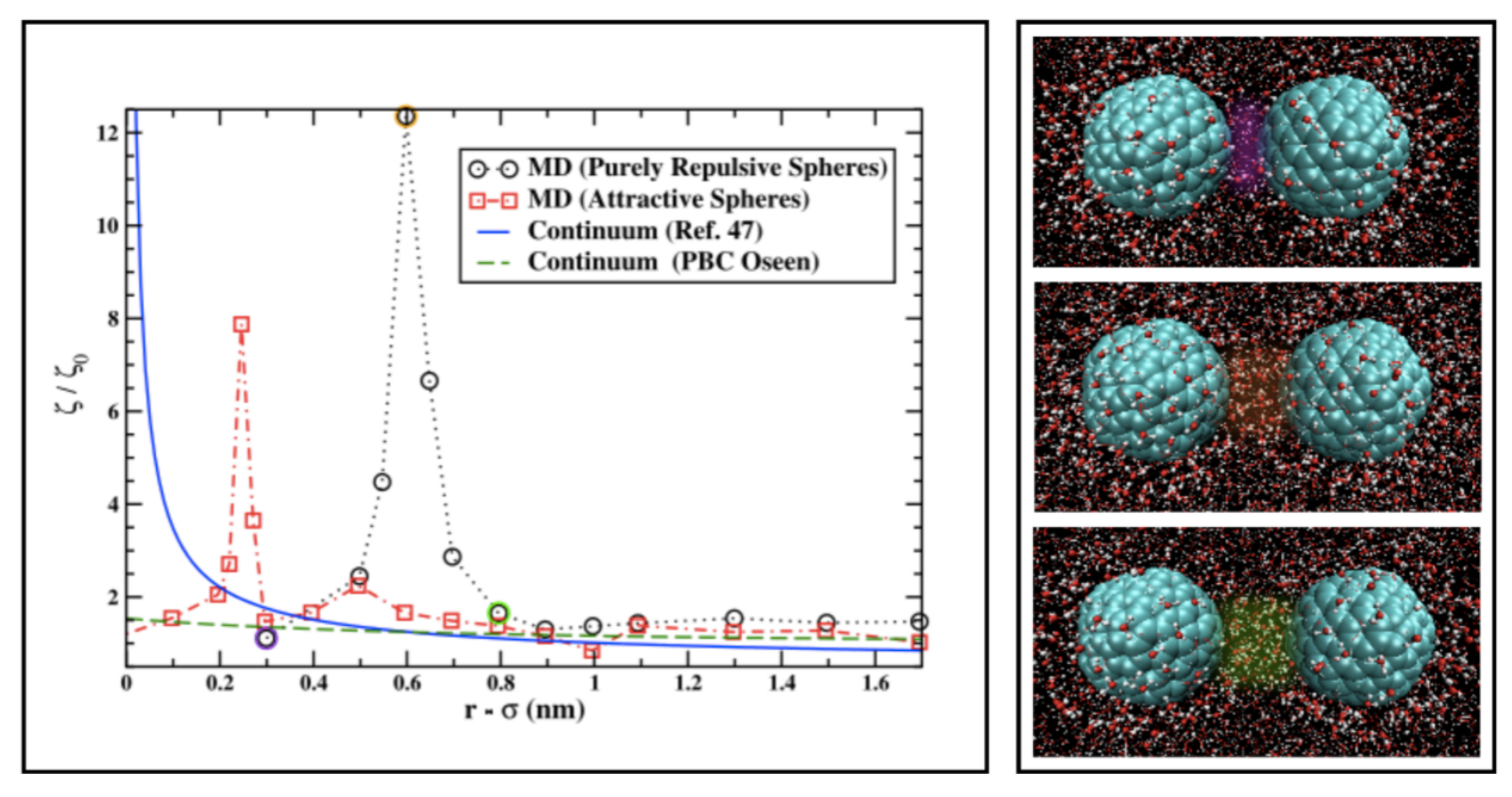}
\caption{The friction coefficient is plotted as a function of the
  distance between two C240 molecules when the WCA potential and LJ
  potential is utilized to model solvent-solute interactions (black
  dotted line with circles and red dashed line with squares,
  respectively).  The two-body (continuum) hydrodynamics result of
  Ref. \onlinecite{Jeffrey:1984tu} (solid blue line) and the periodic
  Oseen tensor (dashed green line) are plotted against the
  molecular-scale computation.  The three right-hand panels depict
  snapshots at separations (increasing from top to bottom) where, in
  the case of the WCA spheres, the inter-solute region is dry and the
  friction coefficient is low (violet circle), at the dewetting
  transition where the friction peaks coefficient (orange circle), and
  for large distances where the friction coefficient decays towards
  the baseline result (green circle).  Water molecules within $0.6$ nm
  of the fullerene surface are depicted by a ball-and-stick
  representation.  All others are represented as lines. The snapshots
  were visualized with VMD.~\cite{vmd96} } \label{fig:panels}
\end{center}
\end{figure*}

The friction
coefficient on the relative coordinate has been computed as a function
of the distance between two spherical bodies where the solute-solvent
interaction is either attractive or purely repulsive.  We find starkly
different behavior in each case, as will be discussed below.  In both
cases, there is significant deviation from continuum hydrodynamic
predictions at small separations.  In Fig. \ref{fig:panels} we plot the
relative friction coefficient as a function of the sphere separation
$r-\sigma$ for C240 where solute-solvent interactions are purely
repulsive or include attraction. Two expressions from continuum
hydrodynamics, the two-body result valid at small separations and the
result of the Oseen tensor corrected for periodic boundary conditions
that is valid at long range, are also
plotted.~\cite{Jeffrey:1984tu,Hasimoto:1959p589,Lindbo:2010ha,Nakayama:2008p587}
One can see that although the friction coefficient for $r-\sigma > 1
\text{ nm}$, approaches the long-range continuum result, the
short-range value does not diverge at the point of contact as
predicted by the expression of Ref. \onlinecite{Jeffrey:1984tu}.
Instead, the frictional profile exhibits distinctive features of
molecular-scale effects.  For purely repulsive bodies, the friction
coefficient peaks at a distance that corresponds to the critical
distance for drying (see Sec. \ref{sec:repuls}).  This distance increases
with solute length scale~\cite{Margulis:2003wo} and such drying
phenomena is not encompassed by continuum hydrodynamics.  In the case
of bodies that strongly attract solvent, drying does not occur, and
water is expelled by steric repulsion.  This result is dependent upon
the chosen value of $\epsilon_{SS}$.  More weakly attractive spheres
can exhibit dewetting.  As discussed in Sec. \ref{sec:attr}, distinctive
molecular-scale behavior in which the friction coefficient exhibits
signatures of solvent layering is observed as solvent is expelled from
the inter-solute region.  Furthermore, we note that the value of the
long-range molecular scale results plotted in Fig. \ref{fig:panels} appear
to approach the description given by the periodic Oseen
tensor. However, data points at larger separations would be necessary
to make more definitive comparisons.

The occupancy and fluctuations of solvent molecules in a particular
region of space may be monitored in a probe volume. This technique is
a standard tool to study the hydrophobic effect (see
e.g. Ref. \onlinecite{Jamadagni:2010bc}).  Nested cylindrical probe
volumes are presently employed to study the water density and
fluctuations in the inter-solute region.  The probe cylinder is
partitioned into an outer shell and an inner-tube.  Due to the
curvature of the hydrophobic surfaces, the density may be
significantly different in the two shells, a finding that is
particularly true in the case of the fullerenes which attract solvent.
A diagram of the boundaries of the probe volume is sketched out in
panel (c) of Fig. \ref{fig:mixed}.  The length of the cylinder is
determined by the separation of the surfaces of bodies. The diameter
of the cylinder is formed by a line tangent to the effective spherical
surface of the fullerene and opposite an angle of $90^\circ$.  The
inner-tube is bounded in the same fashion except the angle opposite
the tangent segment is $60^\circ$.

\subsection{Purely repulsive solute-solvent
  interactions} \label{sec:repuls} 
\begin{figure*}
\begin{center}
\includegraphics[scale=0.55]{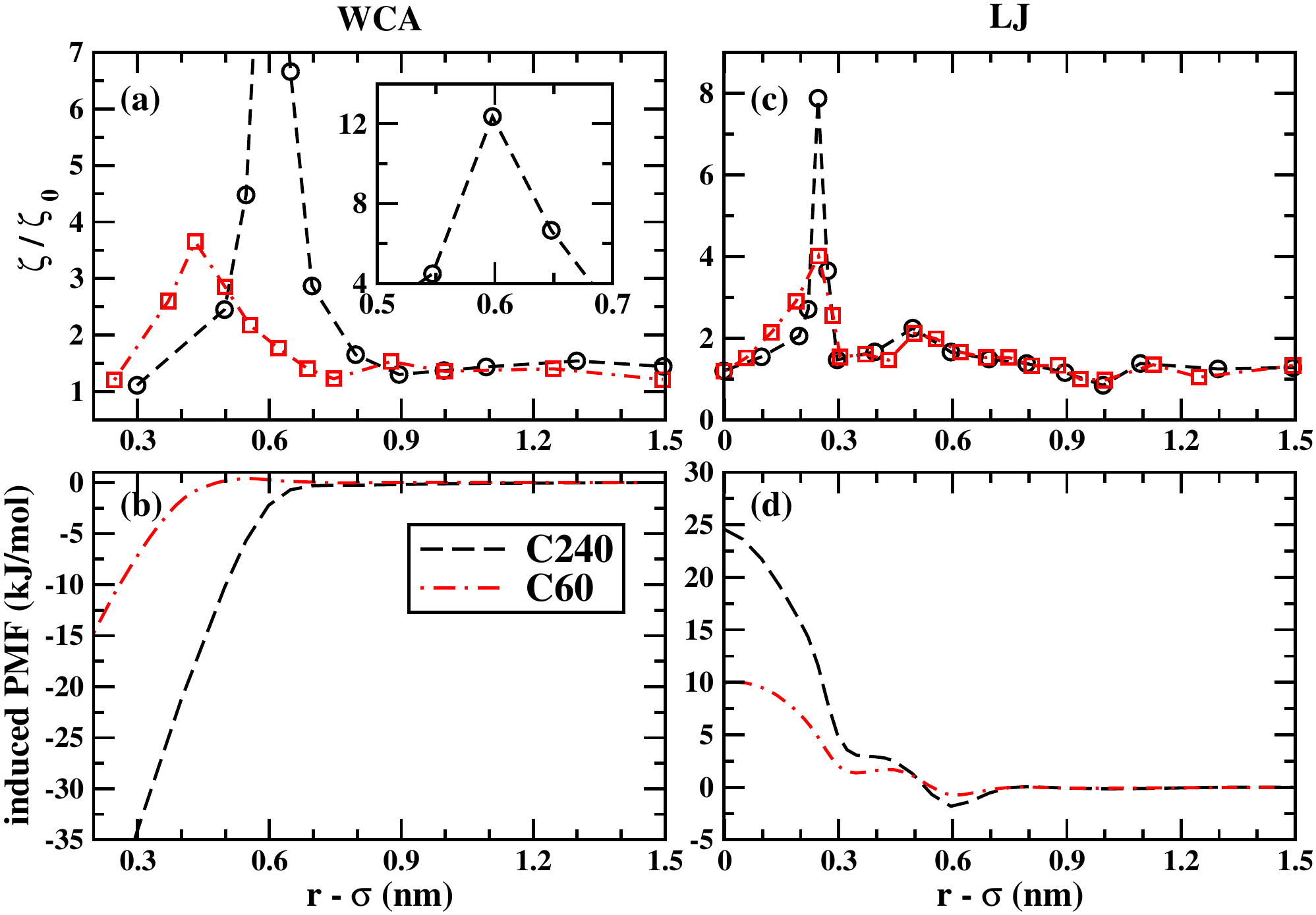}
\caption{The spatial friction coefficient (panel a) and
  solvent-induced potential of mean force (panel b) as two C60 (red
  curve) and two C240 (black curve) approach each other when the
  solute-solvent interactions are purely repulsive.  As expected,
  there is a greater driving force for assembly with increasing solute
  length-scale. Panels (c) and (d) depict, respectively, the spatial
  friction and solvent-induced potential of mean force as two C60 (red
  curve) and two C240 (black curve) approach each other when the
  solute-solvent interactions include attraction. In this case, no
  drying is observed and there is a barrier for desolvation of two
  water layers as the two bodies approach each other.  The friction
  coefficient is not given at the point of contact for the WCA spheres
  due to the substantial uncertainty in the correlations of the force
  fluctuations when the mean force is large. } \label{fig:fricall}
\end{center}
\end{figure*}

\begin{figure*}
\begin{center}
\includegraphics[scale=0.55]{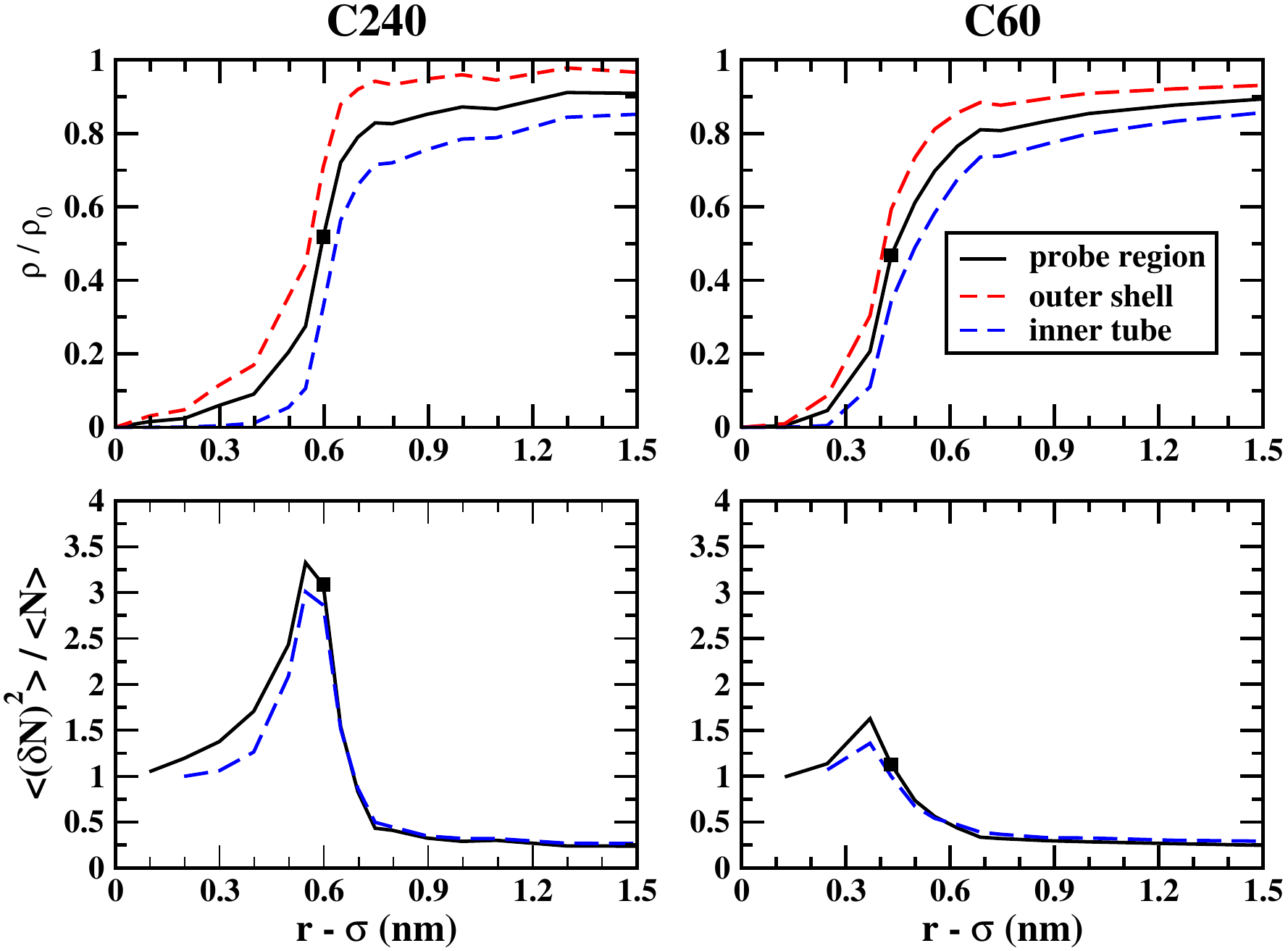}
\caption{The relative density of water (top row) and the ratio of the
variance to the average of the number of water molecules (bottom row)
in the probe volume is plotted for C240 (right column) and C60 (left
column) when the solute-solvent interaction is purely repulsive.  The
probe region is divided into an outer shell (red dashed line) and an
inter-tube (blue dashed line).  The squares indicate the separation
at which the friction coefficient is maximum.} \label{fig:watwca}
\end{center}
\end{figure*}

In this section the results when
only repulsive interactions with the solvent are present is discussed.
The frictional profiles for C60 and C240 are given alongside their
respective solvent-induced potentials of mean force (PMF) in panels
(a) and (b) of Fig. \ref{fig:fricall}.  It can be seen that the hydrophobic
collapse along the relative coordinate is essentially
barrierless. This finding is in agreement with previous simulations of
the assembly of two hydrophobic bodies in both coarse-grained and
explicit solvent
models.~\cite{Margulis:2003wo,Huang:2005ho,Willard:2008p84} The
density and fluctuations of water in the inter-solute regions are
depicted in Fig. \ref{fig:watwca}.  The onset of drying is at larger
distances with increasing solute size, an observation which is in
agreement with prior work.~\cite{Margulis:2003wo} The drying
transition for the approach of two C240 molecules occurs when
approximately two layers of water would be present in the inter-solute
region if only steric effects were considered.  Upon inspection of
Fig. \ref{fig:watwca}, it is apparent that the ratio of density of solvent
to its bulk value is less than unity even at large separations when
the region is wet, an indication that water is depleted near the
surface of the hydrophobe, a result in agreement with the
pair-correlation function shown in Sec. \ref{sec:single}.

The fluctuations of water in the probe volume are plotted in the lower
panels of Fig. \ref{fig:watwca}.  The chosen metric of fluctuations $\left<
\delta N^2 \right> / \left<N\right>$ for a bulk fluid is proportional
to the isothermal compressibility in the limit of large probe volumes,
which naturally does not presently apply.  The fluctuations peak as
the drying transition occurs.  These findings are consistent with the
understanding of dewetting phenomena that have been put forth in
previous studies.~\cite{Margulis:2003wo}

\begin{figure}
\begin{center}
\includegraphics[scale=0.6]{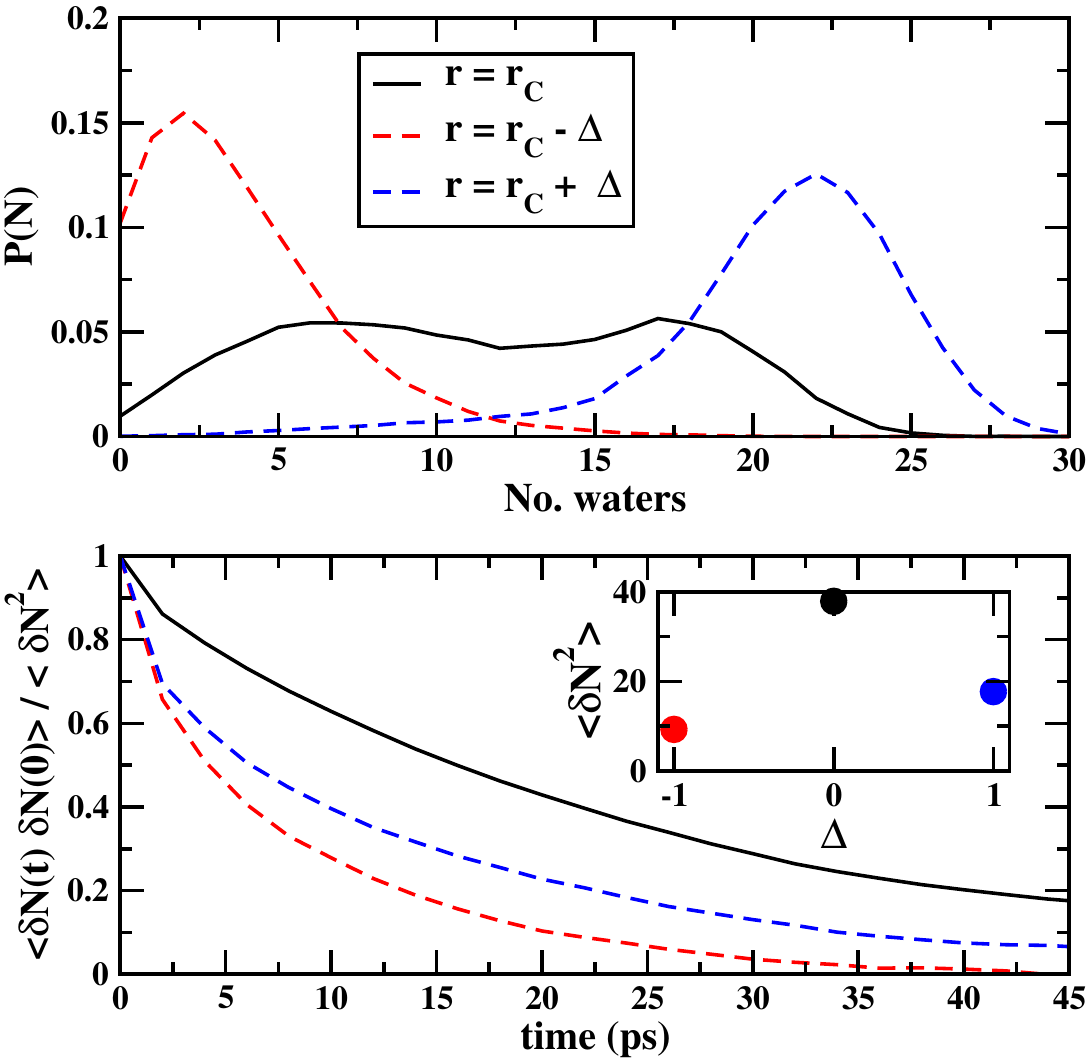}
\caption{(Top panel) The distribution of number of water molecules
present in the probe volume at the critical distance for drying, $r_c
- \sigma = 0.6$ nm (black curve) and at $r=r_c \pm \Delta $ where
$\Delta=0.1$ nm (red and blue dashed lines).  (Bottom panel) At the
critical distance there is a slow relaxation observed in the
autocorrelation function of the number of molecules in the probe
volume as compared to states that are primarily dry or wet. The inset
depicts the variance of the number of waters in the probe volume as a
function of $\Delta$.  All data in this figure pertains to the C240
system when solute-solvent interactions are purely
repulsive.} \label{fig:dewet}
\end{center}
\end{figure}

In order to gain an understanding of how molecular-scale effects
manifest themselves in the spatially dependent friction coefficient,
we now show how the value of the friction coefficient relates to
drying phenomena.  The frictional profile is depicted in panel (a) of
Fig. \ref{fig:fricall} and is strongly peaked at the dewetting transition.
For larger solutes such as C240, this may be defined as the separation
where the inter-solute region fluctuates between ``wet'' and ``dry''
states.~\cite{Margulis:2003wo} The histogram of the number of waters
in the probe volume at $r_c$ is shown in the top panel of
Fig. \ref{fig:dewet}.  There is a bimodal distribution where the difference
between maxima is $\approx 10$ water molecules.  The distributions at
$r = r_c \pm 0.1$ nm are also shown, and at these separations the
inter-solute region is predominantly wet or dry.  Although drying
occurs as two C60 fullerenes approach each other, a bimodal
distribution as exhibited in Fig. \ref{fig:dewet} is less apparent.  In
addition, the critical separation is near the inflection point of the
density curve and the maximum of the solvent fluctuations as
determined by the chosen metric. The point at which the friction
coefficient peaks is marked in the graphs plotted in Fig. \ref{fig:watwca}.

In addition to large static fluctuations (see Fig. \ref{fig:watwca}), slow
relaxation times are also indicative of the drying transition. The
autocorrelation function of the fluctuations in the number of water
molecules in the probe volume,$\left< \delta N(t) \delta N(0)\right>$,
may be decomposed into a product of the variance $ \left<\delta
N^2\right>$ and some time dependent function $h(t)$ that characterizes
the relaxation,
\begin{align} \left< \delta N(t) \delta N(0) \right> &= \left<\delta
N^2\right> h(t) \; .
\end{align} The normalized time correlation function $h(t)$ and the
variance are plotted in the bottom panel of Fig. \ref{fig:dewet}.  One can
see that, in addition to a large variance, slow relaxation times are
associated with collective water dynamics at the critical distance for
drying.  In this way both static and dynamic terms contribute to the
solvent behavior that engenders a large friction coefficient.  The
relaxation of the correlation function is markedly faster when $r=r_c
\pm 0.1$ nm, and the friction coefficient at these separations is
lower by a factor of $\approx 4$.  The computation of the hydrodynamic
interactions is a probe of the momentum relaxation of the solvent (see
Appendix) and slow collective motion of water molecules exiting and
entering the inter-solute region gives rise to, in part, the large
friction that is observed.  However, as the time scale of solvent
relaxation approaches that of the solute, the separation of time
scales essential to the (Markovian) Langevin description will not
hold.  Such possibilities will be discussed in Sec. \ref{sec:conc}.

\begin{figure}
\begin{center}
\includegraphics[scale=0.45]{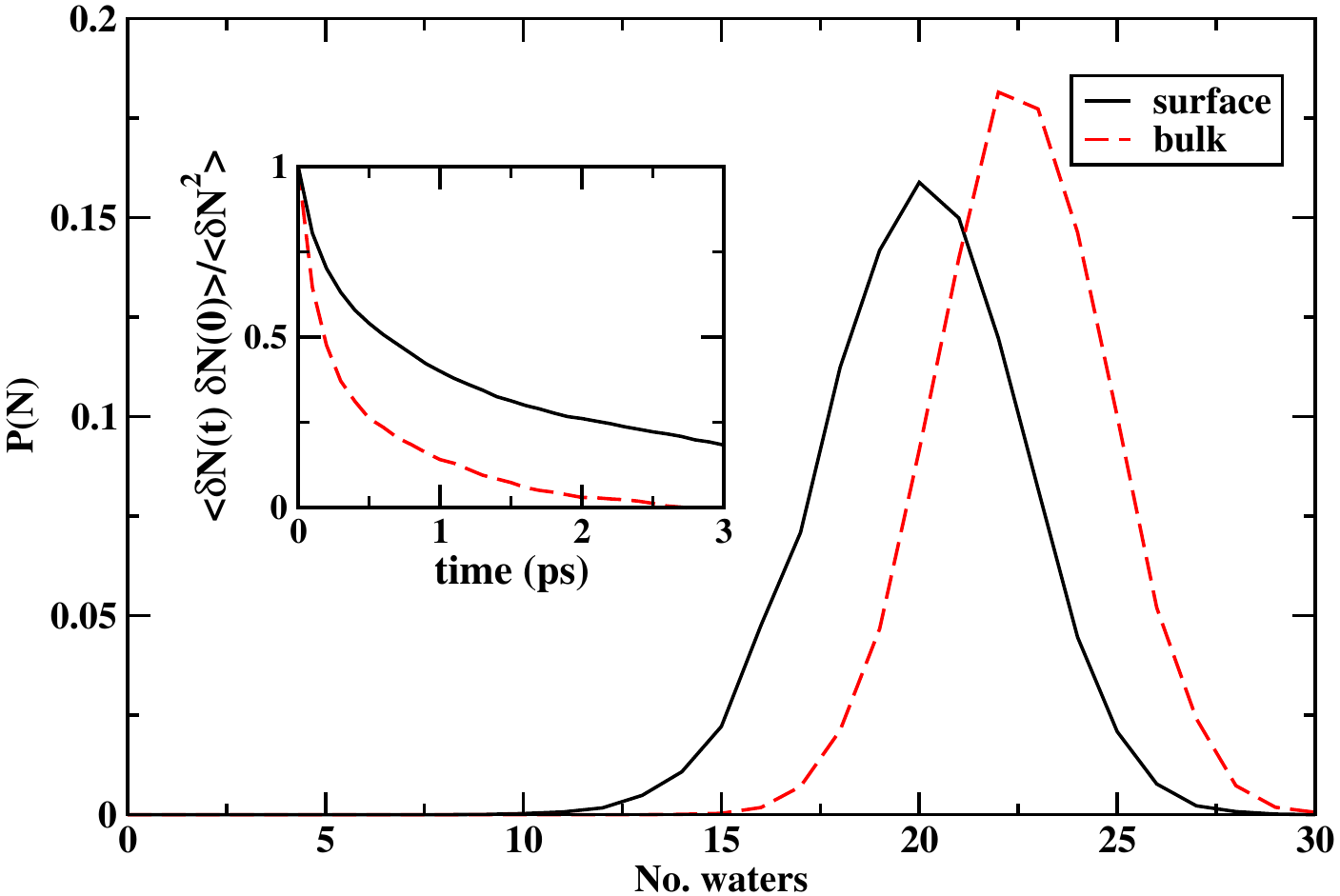}
\caption{The distribution of water molecules in a probe volume of
length $r_c-\sigma$ near the surface of a single C240 molecule (solid
black curve) and in the bulk (dashed red curve) when the
solute-solvent potential is purely repulsive.  The inset depicts the
autocorrelation function of the number of waters in the probe volume
in the respective environments.} \label{fig:dewet2}
\end{center}
\end{figure}

The distribution of water molecules in the same size probe volume used
at $r=r_c$ is given near the surface of a single C240 and in the bulk
in Fig. \ref{fig:dewet2}.  The depletion of water density at the surface is
indicated by the differences in the bulk and single surface
distributions.  Furthermore, the enhancement in fluctuations near the
surface is indicated by the broader distribution.  These findings are
in broad agreement with prior work.~\cite{godawat09,patel10} The
faster relaxation of the number correlation function is seen in the
bulk and near the surface.  This indicates not only the slow
collective motions associated with the bi-stability of the drying
transition, but also the more general effect of confinement between
the two solutes lengthens the relaxation times in the probe volume.

\subsection{Attractive solute-solvent interactions} \label{sec:attr}
\begin{figure*}
\begin{center}
\includegraphics[scale=0.55]{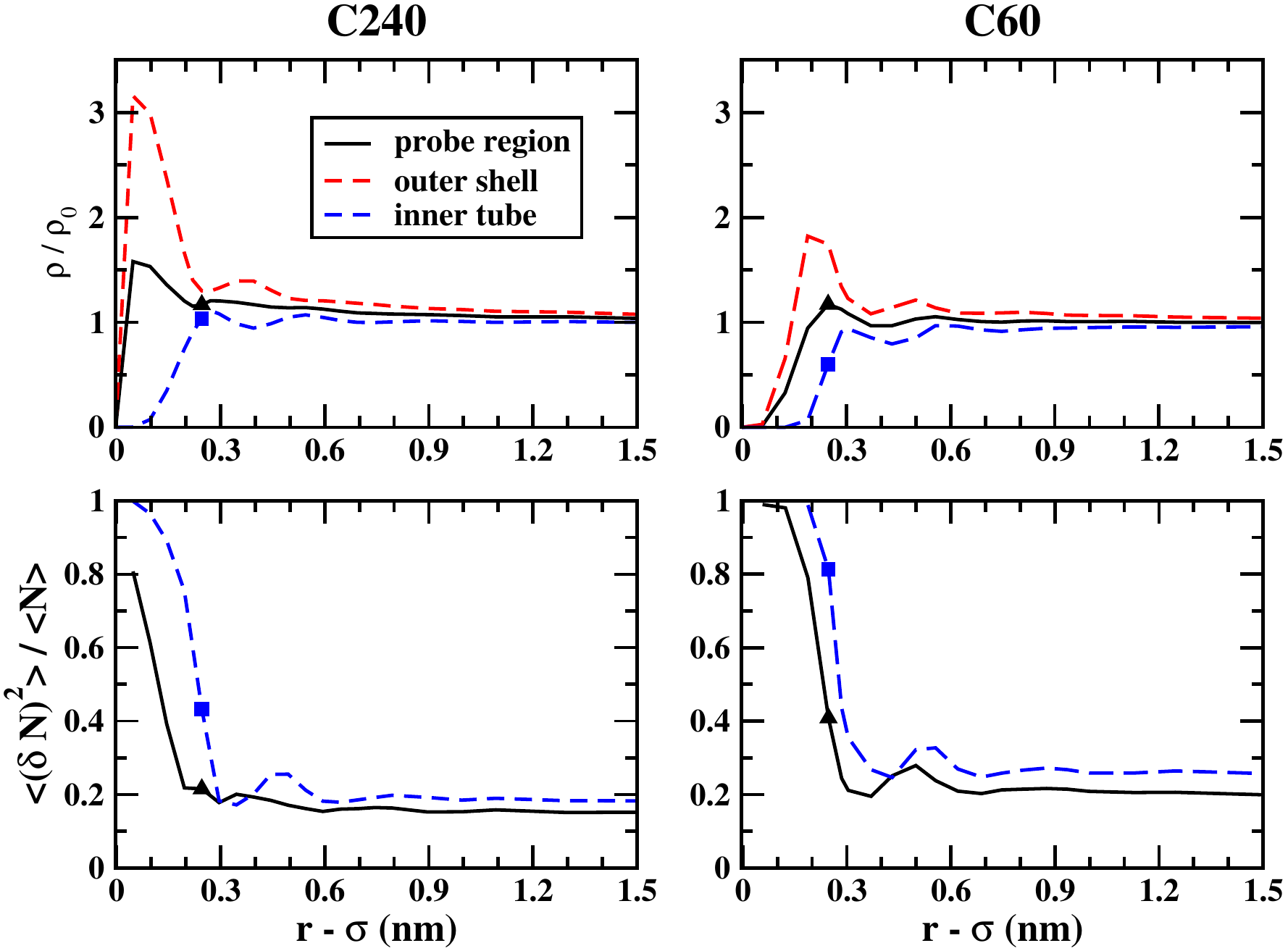}
\caption{The relative density of water (top row) and the ratio of the
variance to the average of the number of water molecules (bottom row)
in the probe volume is plotted for C240 (left column) and C60 (right
column) when the solute-solvent interaction includes attraction. The
probe region is divided into an outer shell (red dashed line) and an
inner-tube (blue dashed line).  Triangles and squares denote the
separation at which the friction coefficient is maximum. } \label{fig:watlj}
\end{center}
\end{figure*}

Given the high density of carbon sites and the value of
$\epsilon_\text{SS}$ chosen for the Lennard-Jones potential, the
solutes possess a sizable affinity for water when attractive
interactions are included.  The solvent density is plotted in
Fig. \ref{fig:watlj}.  The curvature of the surface has a significant
impact on water density and fluctuations.  For small separations where
solvent in the inter-solute region can experience attraction from two
bodies, the density is enhanced in areas in the outer probe region
where steric repulsion does not dominate.  High densities are
accompanied by large fluctuations of the number of water molecules in
the outer shell.  Due to surface curvature, water is first squeezed
out from the inner probe volume as the density decreases in this
region as the second layer is vacated, and subsequently, the density
falls to zero as the final layer of water is removed.  In analyzing
the trends in the friction coefficient and potential of mean force,
the inner probe volume is more indicative of solvent layering.

The induced potential of mean force depicted in the panel (d) of
Fig. \ref{fig:fricall} exhibits minima at separations where one and two
water layers are stable at the narrowest portions of the inter-solute
region, as indicated by monitoring the inner probe volume.  As the
spheres come into contact there is a maximum in the solvent induced
PMF.  Prior calculations of the PMF for a pair of C60 molecules are in
qualitative agreement.~\cite{Makowski:2009p61} Prior computations of
the PMF as two graphene plates approach exhibit barriers for the
removal of each solvent layer,~\cite{Zangi:2011p581} although the
curvature of the spheres gives rise to different behavior than plates
at small separations.  Namely, as the spheres are curved, the
inter-solute region never fully desolvates and the induced mean force
opposes assembly until the point of contact.

Upon examination of the spatial dependence of the friction coefficient
(panel (c) of Fig. \ref{fig:fricall}), evidence of layering is exhibited by
the non-monotonic behavior of the friction coefficient for small
separations $r-\sigma<0.5$ nm.  The friction coefficient peaks as the
PMF rises and the second layer of solvent is removed.  At smaller
separations, the friction coefficient falls at the minimum of both the
induced PMF and solvent fluctuations ($r-\sigma=0.3$ nm), and then as
the final water layer vacates the inter-solute region both the PMF and
friction coefficient rise sharply, with the size of the latter abating
rapidly as solvent density in the inner region decreases.  These
trends are also witnessed in the water statistics.  The point at which
the friction coefficient is maximum corresponds to a point where the
density is near a maximum, and the fluctuations are large and
increasing.  The values of the density and fluctuation at the
separation of maximum friction coefficient are marked on the curves
presented in Fig. \ref{fig:watlj}.  In the case of C240, these trends are
more clearly observed upon study of the smaller probe volume.  It
stands to reason that in the case of attractive spheres, the friction 
exhibits some direct relation to both the density and fluctuations of
solvent in the inter-solute region.

The spatial dependence of the friction coefficient is in stark contrast to the
predictions of continuum hydrodynamics in the short-range limit, where
the friction coefficient is predicted to monotonically increase and diverge as the
bodies are brought into contact.  This discrepancy is engendered by
the layering and granularity of liquids over small length scales.  The
increase in friction coefficient may have its molecular origins in the confined
rattling in the inter-solute region as the water is squeezed
out,~\cite{Bocquet:1997p103} as is indicated by the large fluctuations
in the probe volume (Fig. \ref{fig:watlj}). It is also observed that the
spatial friction coefficient plotted in panel (c) of Fig. \ref{fig:fricall} scales with
particle size, that is the curves for C60 and C240 nearly superimpose
upon each other when the difference in effective molecular radius is
taken into account.

\subsection{Diffusion controlled rate} 
\label{sec:rate} 

\begin{figure*}
\begin{center}
\includegraphics[scale=0.50]{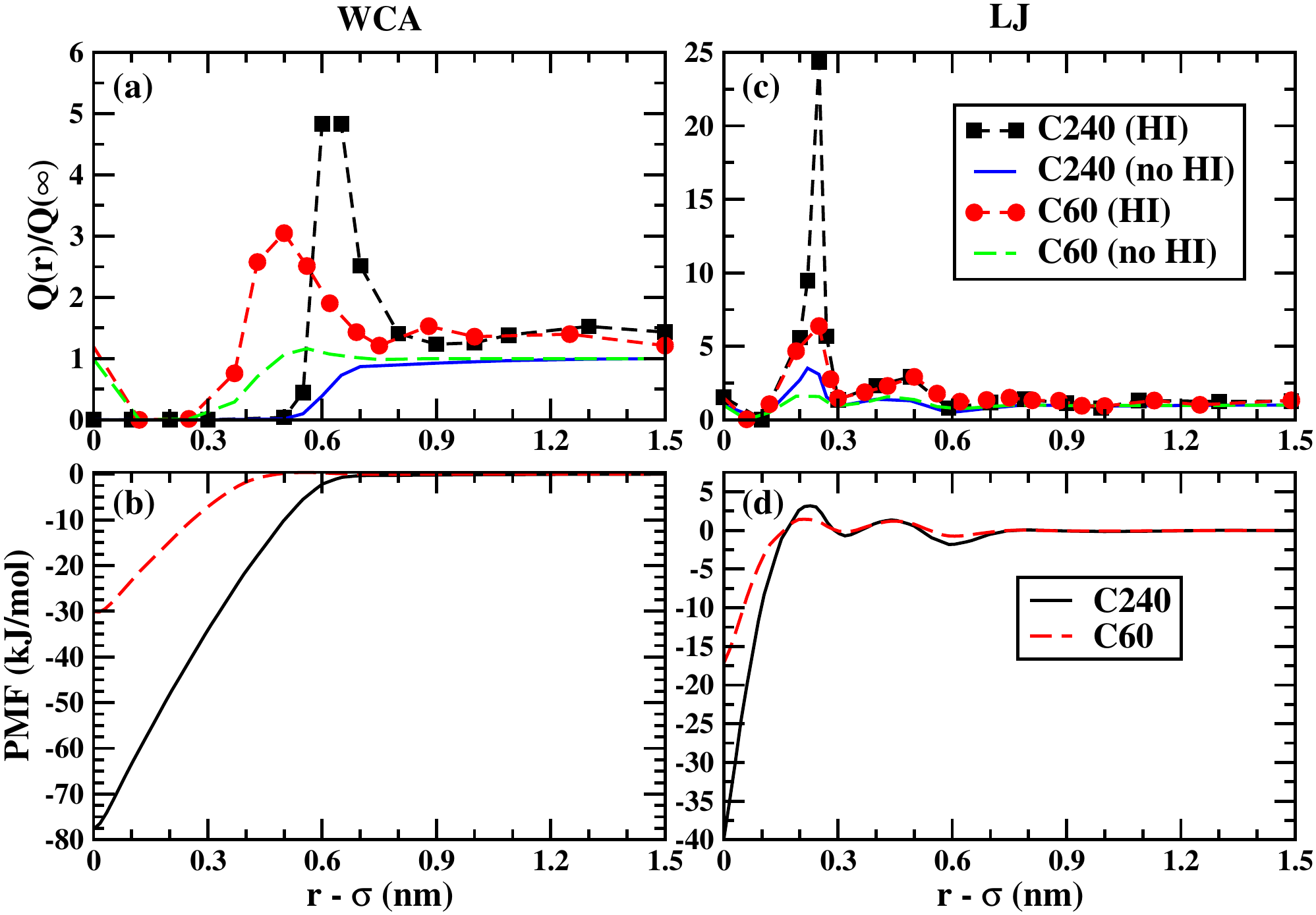}
\caption{The ratio of the friction coefficient to pair-correlation function,
$Q(r)$ is plotted with and without the inclusion of hydrodynamic
interactions (HI) as two C240 
(black curves that are dotted with squares and dashed, respectively) 
and two C60 (red curves that are dotted with circles and dashed, respectively) 
molecules approach when the solute-solvent
interactions include attraction (panel a) and for the case when the
solute-solvent interaction are purely repulsive (panel
c). $Q(\infty)=\zeta_0$ is defined as the value of $Q$ for large
separations when the potential of mean force and hydrodynamic
interactions are negligible. The potential of mean force including the
direct interactions between a pair of C240 and C60 molecules with
solute-solvent interactions that are purely repulsive (panel b) and
include attraction (panel d) is also depicted.} \label{fig:allrate}
\end{center}
\end{figure*}

In the high friction limit,
a radial Smoluchowski equation describes the motion along the relative
coordinate, $r$.  The diffusion-controlled rate of reaction can be
computed from the following
expression~\cite{emeis70,deutch73,wolynes76,Northrup:1979wr,calef83,Straub:1990p65}:
\begin{align} k^{-1} = \frac{1}{4\pi k_b T} \int\limits_\sigma^\infty
\mathrm{d} r \; \frac{Q(r)}{r^2} \label{eqn:rate}
\end{align} where $\sigma$ is the contact diameter of the Brownian
bodies and,
\begin{align} Q(r) = \frac{\zeta(r)}{g(r)} \; ,
\end{align} is the ratio of the friction coefficient to the pair-correlation
function $g(r)=e^{-\mathcal{W}(r)/k_bT}$, where $\mathcal{W}(r)$ is
the total potential of mean force that includes contributions from
both direct and (induced) solvent-mediated interactions. The full PMF
is plotted in panels (b) and (d) of Fig. \ref{fig:allrate} and for C60 is
in good agreement with prior work.~\cite{Makowski:2009p61} The
function Q(r) is plotted in panels (a) and (c) of Fig. \ref{fig:allrate}
for all combinations of fullerene sizes and interaction potentials
presently considered.  It is compared with its value if hydrodynamic
interactions are neglected, that is when $\zeta(r) = \zeta_0$, where
$\zeta_0$ is one half the friction coefficient on the single body under periodic
conditions (see Sec. \ref{sec:single}).

In the case of purely repulsive solute-solvent interaction, Q(r) has a
markedly different form that depends on whether or not hydrodynamic
interactions are included.  In particular, due to the fact that the
PMF is virtually barrierless when solute-solvent interactions are
purely repulsive, hydrodynamic interactions give rise to a maximum in
Q(r), as exhibited in panel (a) of Fig. \ref{fig:allrate}.  In the case of
bodies that attract solvent, Q(r) is also larger when hydrodynamic
interactions are included.  However in contrast to ideal hydrophobes,
the inclusion of spatially dependent friction coefficient tends to ``reinforce''
the extrema engendered by the PMF.  Consistent with the frictional
profiles presented in Fig. \ref{fig:fricall}, the impact of hydrodynamic
interactions is greatest at separations where drying or desolvation
occurs. The function $Q(r)$ is found to increase at small separations,
$r-\sigma< 0.1 \; \text{nm}$, as direct repulsive interactions
dominate in this region.

In order to evaluate Eqn. \ref{eqn:rate}, the spatial friction coefficient
is approximated for large separations via a switching
function~\cite{Morrone:2010p363} that smoothly reduces its value to a
plateau of $\zeta_0$ at a distance $0.5$ nm further than the farthest
separation for which the friction coefficient was computed.  Although
the computed rate constant exhibits weak dependence on the choice of
cutoff, this does not alter the observed trends.  We note that
$\zeta_0$ depends on the periodic boundary conditions adopted (see
Sec. \ref{sec:single}), and, moreover, due to the small relative
separations studied in our work, our calculations do not capture the
long-range decay of the hydrodynamic interactions.  Thus the rate
constant for association presently computed accurately reflects the
short-range hydrodynamic interaction at the molecular scale but not
the contribution of the long range hydrodynamic interactions.

If both the potential of mean force and hydrodynamic interactions are
neglected the diffusion controlled rate constant reduces to the familiar
expression,
\begin{align} k_0 &= \frac{4\pi k_b T \sigma}{\zeta_0} \;.
\end{align} The ratio of the rates to $k_0$ computed by means of
Eqn. \ref{eqn:rate} with and without the inclusion of hydrodynamic
interactions or the potential of mean force are given in
Table \ref{tab:rate}.  The values of $k_0$ are nearly independent of
molecular diameter, which is a consequence of the Stokes-Einstein
relation and the direct proportionality of friction coefficient to
solute diameter.  It can be seen that if the potential of mean force
is considered and hydrodynamic interactions are not included (that is
$\zeta(r)=\zeta_0$), then the rate of ``ideal'' hydrophobic assembly
(when the solute only excludes volume) is increased.  In the case of
the attractive solutes, there are barriers to assembly and an
attractive basin at short range that serve to only modestly impact the
rate. The inclusion of hydrodynamic interactions reduces the diffusion
controlled rate by approximately $30-40\%$ from its respective value
across all systems presently considered. If the potential of mean
force (both direct and solvent-induced) is not included in the
computation of the rate constant, then one can clearly see that
hydrodynamic interactions reduce the rate of association almost
uniformly across the four systems studied, this despite the different
physics in their separate behaviors (see Fig. \ref{fig:allrate}).
This is probably due to the fact that the main quantitative impact of
molecular-scale hydrodynamics is to increase the value of $Q(r)$ at
separations slightly larger than required for desolvation of the
inter-solute region.

Continuum hydrodynamic theory has been utilized to compute the rate of
assembly of two bodies.~\cite{deutch73,wolynes76,sun07,elcock10} As is
well known, the strong divergence of the friction coefficient
exhibited by the short-range lubrication expression utilized in
Stokesian dynamics and plotted in Fig. \ref{fig:panels} prohibits contact
and thus the computed rate constant from Eqn. \ref{eqn:rate} would be
zero.~\cite{Jeffrey:1984tu,Brady:1988tz,Brady:1988p970} The
lubrication limit for the case of slip boundary conditions as derived
by Wolynes and Deutch~\cite{wolynes76} is integrable and hydrodynamic
interactions were found to decrease the rate constant by $\approx
30\%$.  Recently, hydrodynamics were included by means of the
Rotne-Prager tensor in the evaluation of protein-protein association
rates and found to decrease the rate of association by
$\approx35-80\%$.~\cite{elcock10} Although these results generally
appear in agreement with our work, as noted above, the present
estimates only adequately gauge the impact of short-range hydrodynamic
interactions and therefore do not readily lend themselves to detailed
comparisons.

As stated above, if the spatial dependence of the friction coefficient is
neglected, the rate constant for reaction is larger than $k_0$ for
ideal hydrophobes, as the solvent induced potential of mean force is
attractive.  This rate constant is reduced by hydrodynamic interactions which,
in some sense, capture the timescale associated with dewetting
fluctuations (see Sec. \ref{sec:repuls}).  As drying fluctuations are
relatively slow (see Fig. \ref{fig:dewet}) this raises the question of
whether or not the separation of timescales within the Brownian limit
is good description of systems near a drying transition.

\begin{table*}
\begin{center}
\begin{tabular}{||c|c|c|c|c|c||} \hline & & $k/k_0$ (HI) & $k/k_0$ (No
HI) & $k/k_0$ (No PMF) & $k_0$ ( $10^{10}$ M$^{-1}$ s$^{-1}$ )\\ \hline C60 & LJ & 0.66 &
1.02 & 0.68 & 1.16 \\ C240 & LJ & 0.60 & 0.95 & 0.74 &1.10 \\ C60 & WCA & 0.93 &
1.27 & 0.72 & 1.54 \\ C240 & WCA & 0.99 & 1.37 & 0.64 & 1.51 \\ \hline
\end{tabular}
\caption{Table of the ratio of the diffusion controlled rate constants with and
  without the inclusion of hydrodynamic interactions or the potential
  of mean force to the quantity $k_0=4\pi k_b T
  \sigma/\zeta_0$} \label{tab:rate}
\end{center}
\end{table*}

\section{Discussion and conclusions} \label{sec:conc} 
The spatial
dependence of the friction coefficient in the Brownian limit has been
computed as two non-polar bodies come into contact.  The system is
analyzed within the framework of Brownian dynamics and hydrodynamic
interactions are included via explicit molecular dynamics
simulation. We find the friction coefficient deviates from continuum
hydrodynamic predictions at small separations and dramatically depends
on the nature of solute-solvent interactions.  For purely repulsive
spheres, we find that the friction coefficient peaks at the critical
distance for dewetting and decreases as the inter-solute region dries.
For attractive solutes water is expelled by steric repulsion and the
effects of solvent layering are apparent in the non-monotonic
dependence of the friction coefficient on separation.  The variation
of friction coefficient with separation is due solvent occupation and
fluctuations in the inter-solute region.  Large solvent fluctuations
and slow relaxation times are associated with increasing friction
coefficient, and in this way both static and dynamical
solvent-mediated effects contribute to the frictional force. Under
certain circumstance slow solvent fluctuations in the near drying
regime may give rise to non-Markovian effects.

In their study of hydrophobic assembly, Willard and Chandler found
that not only the relative separation, but also solvent degrees of
freedom, namely the occupancy of the inter-solute region, are
necessary to characterize the process.~\cite{Willard:2008p84}  Such
behavior has also been observed in the assembly of hydrophobic
plates.~\cite{Margulis:2003wo,Huang:2005ho} As our analysis is in the
framework of Brownian dynamics, only solute degrees of freedom are
explicitly treated in the rate calculation presented in
Sec. \ref{sec:rate}.  However, solvent fluctuations engender the peak in
the spatially dependent friction coefficient observed at the dewetting
transition.  In this way, drying phenomena manifests itself when
molecular-scale hydrodynamic interactions are included in the Brownian
limit.

The Brownian description is valid if the motion of the solute occurs
on a much slower time scale than the solvent dynamics.  We find that
relatively slow solvent motions are present at the critical separation
for dewetting, which raises questions about the validity of the
Markovian assumption.  We found the correlation time for water
fluctuations in the inter-solute region at the critical separation to
be $\approx 25$ ps for C240.  (see Fig. \ref{fig:dewet}). Prior work has
shown that the timescale for hydrophobic assembly depends on not only
the size of the solutes but also on the nature of the
interactions,~\cite{Margulis:2003wo,Huang:2005ho} and the time scale
for solvent fluctuations can be much longer than 25 ps.  In the case
where solvent fluctuations cannot be considered fast with respect to
solute diffusion, one may consider alternative formulations that
include dynamical disorder~\cite{Zwanzig:1990uo}, and the use of
techniques for extracting the spatially dependent memory kernel within
the framework of the generalized Langevin
equation.~\cite{Berne:1970p90,Straub:1987p76,Straub:1990p65}
Naturally, the time scales will depend upon the size of the solute, and
future studies may be designed to more fully probe this observation.

Recently, hydrodynamic interactions have been shown to significantly
decrease \emph{in vivo} diffusion in cellular
environments.~\cite{Ando:2010p99} As the present computation is
limited to two bodies in a solvent bath, one can only draw limited
parallels to behavior in crowded systems. With this caveat in mind, we
find some correspondence in our study where we find that inclusion of
hydrodynamic interactions leads to a decrease of the diffusion
controlled rate constant for assembly by approximately $30-40\%$. For
barrierless ``ideal'' hydrophobic assembly along the relative
coordinate, hydrodynamic interactions introduce a frictional
``barrier'' that retards the rate of assembly (see Fig. \ref{fig:allrate}),
whereas in the case of attractive solvents the main effect is to
enhance the maxima engendered by the potential of mean force.  In this
way, the interplay between hydrodynamic interactions and the free
energy surface strongly depends on solvent-solute coupling.  Future
work will be aimed at exploring these insights and applying them to
study diffusive phenomena in nanoscopic systems.

\begin{acknowledgements} This research was supported by a grant to
  B.J.B. from the National Science Foundation via Grant No. NSF-CHE-
  0910943. We thank Prof. Kang Kim of the Institute of Molecular
  Science, Okazaki for useful discussions, and Prof. Jeff Skolnick for
  wetting our interest in this problem.
\end{acknowledgements}

\appendix
\section{Calculation of the spatially dependent friction coefficient}
\subsection{Two-body friction tensor} In this work, we study the
spatially dependent friction coefficient along the relative distance between two
bodies, $r=\left|\vec{r}_2 - \vec{r}_1\right|$.  The complete friction coefficient
tensor is related to the motion of all $3N$ degrees of freedom. The
tensor relates the frictional force to the particle velocities.  For
$N=2$ this can be expressed as,
\begin{align} \left( \begin{array}{c} \delta \vec{F}_1 \\ \delta
\vec{F}_2 \end{array} \right) &= \left( \begin{array}{cc}
\mat{\zeta}_{11}(r) & \mat{\zeta}_{12}(r) \\ \mat{\zeta}_{21}(r) &
\mat{\zeta}_{22}(r) \end{array} \right) \left( \begin{array}{c}
\vec{v}_1 \\ \vec{v}_2 \end{array} \right) \; , \label{eq:frictensor}
\end{align} where $\delta \vec{F}$ is the deviation of the force on
the sphere from its mean value.  The friction coefficient tensor $\mat{\zeta}(r)$
has dimension of $6 \times 6$ and may be decomposed into four blocks
that correspond to self and cross interactions between the bodies.
Each submatrix is a diagonal $3 \times 3$ matrix,
\begin{align} \mat{\zeta} = \left( \begin{array}{ccc} \zeta_\| & 0 & 0
\\ 0 & \zeta_\bot & 0 \\ 0 & 0 & \zeta_\bot \end{array} \right) ,
\end{align} if the coordinate system is defined such that one
direction is parallel and two directions are perpendicular to the line
of centers.  The two spheres are identical so by symmetry
$\mat{\zeta}_{11}(r)=\mat{\zeta}_{22}(r)$ and
$\mat{\zeta}_{12}(r)=\mat{\zeta} _{21}(r)$.  The friction coefficient along the
relative coordinate, $r=r_{2,\|} - r_{1,\|}$, may be obtained from
manipulation of Eqn. \ref{eq:frictensor}.  The following relation can then
be extracted,
\begin{align} \zeta_\text{rel}(r) = \frac{1}{2} \left(
\zeta_{11,\|}(r) - \zeta_{12,\|}(r) \right) \; ,
\end{align} and associated to the Langevin equation given by
Eqn. \ref{eqn:rel_lange}.

\begin{figure}
\begin{center}
\includegraphics[scale=0.45]{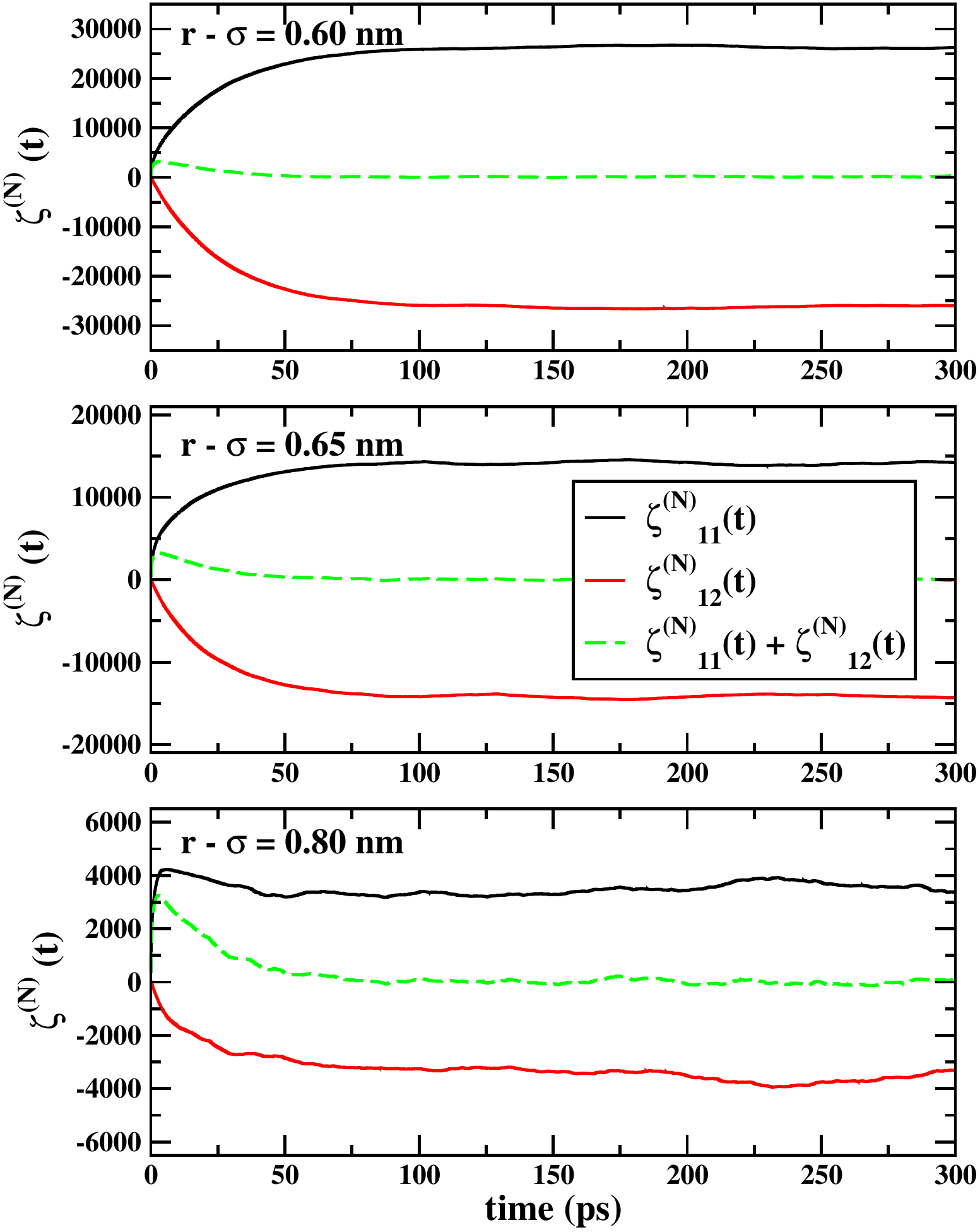}
\caption{The time integral of the self (black curve) and cross (red
curve) force-force autocorrelation functions and the sum of these two
functions (green curve) is plotted as two C240 molecules approach each
other and interact with the solvent purely by means of repulsion. The
top panel, center, and bottom panel depict this quantity at different
separations. One can see that the self and cross autocorrelation
functions plateau at a non-zero value, though their sum decays to
zero, in agreement with the finding of Bocquet \emph{et
al.}~\cite{Bocquet:1997p103} The friction coefficient is plotted in units of
(kj/mol) $\cdot$ (ps/nm$^2$).} \label{fig:plateau}
\end{center}
\end{figure}

 \subsection{Review of techniques relating the friction coefficient to molecular
dynamics simulation} Here we review the techniques developed in
Refs.~\onlinecite{Bocquet:1994p317} and \onlinecite{Bocquet:1997p103} for
the extraction of the Brownian friction coefficient from molecular dynamics
simulations.  First consider a single Brownian sphere of mass, $M$, in
a bath of $N$ solvent molecules of mass, $m$.  This comprises an
isolated system.  Next we consider the limit, $M\rightarrow \infty$,
in which the solute particles are fixed. In this case, the total
solvent momentum is not a conserved quantity. As shown below, the
computation of the friction coefficient is essentially a probe of the total
momentum relaxation.

We begin by relating the rate of change of the total solvent momentum
$P(t)$ to the force acting on the solute $\mathcal{F}(t)$ by means of
Newton's third law.
\begin{align} \mathcal{F}(t) &= -\dot{P}(t) \label{eq:newton}
\end{align} To simplify the notation we only consider one dimension,
although the expressions for three-dimensions are readily obtainable.
Next, consider the integral that in the $t \rightarrow \infty$ and $N
\rightarrow \infty$ limit yields the Green Kubo relation for the
friction coefficient. This can be expressed in terms of the total solvent
momentum:
\begin{align} \zeta(t) &= \frac{1}{ k_b T} \int\limits_0^{t}
\mathrm{d}\tau \left< \mathcal{F}(0) \mathcal{F}(-\tau) \right> \\ &=
- \frac{1}{k_b T} \left< \dot{P}(t) P(0) \right> \label{eq:zetat}
\end{align} Onsager's principle linearly relates the force acting on
the solute to the total solvent momentum in the long time limit:
\begin{align} \mathcal{F}(t) &= \frac{\zeta}{Nm}
P(t) \label{eq:onsager} \\ \dot{P}(t) &= - \frac{\zeta}{Nm}
P(t) \label{eq:diffeq}
\end{align} Where Eqn. \ref{eq:diffeq} arises from combining
Eqn. \ref{eq:newton} and Eqn. \ref{eq:onsager}.  Utilizing Eqn. \ref{eq:zetat}, it
can readily shown that $\zeta(t)$ has the simple form:
\begin{align} \zeta(t) &= \zeta e^{-\zeta t /Nm }
. \label{eq:timezeta}
\end{align} Both the $t \rightarrow \infty$ and $N \rightarrow \infty$
limits must be applied in order calculate the friction coefficient in the linear
response regime (See Eqn. \ref{eqn:greenkubo}).  If the time limit is taken
first, as is necessarily the case when computing the property in a
simulation of finite size, then $\zeta(\infty)\rightarrow 0$.
However, if the thermodynamic limit is taken first
($N\rightarrow \infty$), then a finite and correct value for the
friction coefficient may be recovered.  Prior work has shown that instead of
directly applying the Green-Kubo relation, the friction coefficient may be
recovered by probing the relaxation of $\zeta(t)$.  Presently this was
achieved by an analysis of the Laplace transform of Eqn. \ref{eq:timezeta}
developed in Ref. \onlinecite{Bocquet:1994p317}.

\begin{figure}
\begin{center}
\includegraphics[scale=0.50]{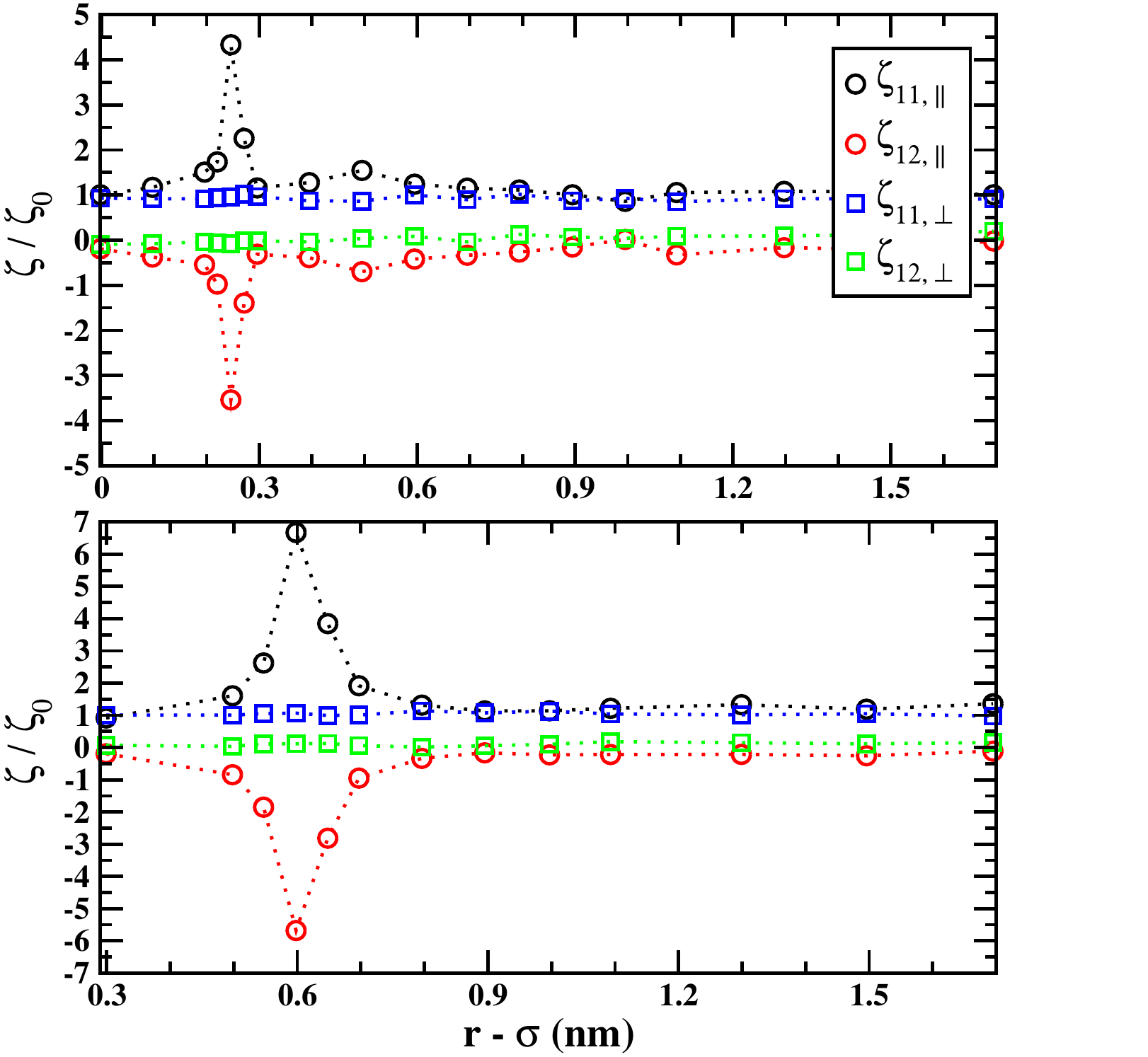}
\caption{The unique components of the friction coefficient tensor as two C240
  molecules approach each other and interact with the solvent with
  attractive (top panel) or purely repulsively (bottom panel) forces.
  The self term parallel (black circles) and perpendicular (blue
  squares) to the line of solute centers is plotted alongside the
  cross friction coefficient in the parallel (red circles) and perpendicular
  (green squares) direction. } \label{fig:tensor}
\end{center}
\end{figure}

In the case of the two body friction coefficient we begin by defining the integral
of the naive MD estimate for each submatrix of the friction coefficient tensor
that one arrives at from the direct application of the Green-Kubo
relation:
\begin{align} \zeta_{11}^{(N)}(t) &= \frac{1}{k_b T} \int\limits_0^t
\mathrm{d} \tau \; \left< \delta F_1 (t) \delta F_1(0)
\right> \label{eq:zeta11} \\ \zeta_{12}^{(N)}(t) &= \frac{1}{k_b T}
\int\limits_0^t \mathrm{d} \tau \; \left< \delta F_1 (t) \delta F_2(0)
\right> \label{eq:zeta22}
\end{align} The superscript $N$ denotes the number of solvent molecules
present in the simulation.  When two Brownian bodies are present, the
solvent momentum relaxation is related to the sum of the fluctuations
of forces on the two spheres, $\dot{P}(t) = - (\delta F_1(t) + \delta
F_2(t))$.  Analogous to case of a single body in a bath reviewed
above, when the long time limit is taken for finite $N$, it was shown
in Ref. \onlinecite{Bocquet:1997p103} that
\begin{align} \zeta_{11}^{(N)}(\infty) + \zeta_{12}^{(N)}(\infty) &= -
\int\limits_0^\infty \mathrm{d}\tau \left< \dot{P}(t) \delta F_1(0)
\right> \nonumber \\ &= 0
\end{align} This spurious result is due to finite size effects and is
demonstrated numerically in Fig. \ref{fig:plateau}.  Although the measured
force-force correlation functions of Eqn. \ref{eq:zeta11} and
Eqn. \ref{eq:zeta22} plateau to non-zero values, their sum does indeed
decay to zero.  Bocquet \emph{et al.} proceed to relate the integrals
of the force-force autocorrelation functions with their proper
$N\rightarrow \infty$ limits.  Details are given in
Ref. \onlinecite{Bocquet:1997p103}.  This analysis yields relations of
the plateaus of the components of $\mat{\zeta}^{(N)}(t)$ to a linear
combination of components of the friction coefficient tensor $\mat{\zeta}$ that
correctly approach the thermodynamic limit.
\begin{align} \frac{1}{2} \left( \mat{\zeta}_{11} - \mat{\zeta}_{12}
\right) = \mat{\zeta}_{11}^{(N)}(\infty) =
-\mat{\zeta}_{12}^{(N)}(\infty) . \label{eq:rel_f}
\end{align} It can be readily seen that the component of the linear
combination parallel to axis of centers can be identified with the
friction coefficient in the relative directions that is reported in
Fig. \ref{fig:fricall}.  In order to obtain the full friction coefficient tensor, one
more relation must be identified,
\begin{widetext}
\begin{align} -\frac{2}{N m k_b T} \left( \mat{\zeta}_{11} +
\mat{\zeta}_{12} \right) t &= \ln \left(
\frac{\mat{\zeta}_{11}^{(N)}(t) +
\mat{\zeta}_{12}^{(N)}(t)}{\mat{\zeta}_{11}^{(N)}(0) +
\mat{\zeta}_{12}^{(N)}(0)} \right) \; ,
\end{align} 
\end{widetext}
which characterizes the relaxation of the sum of the
force-force correlation functions plotted in Fig. \ref{fig:plateau} to
zero.  The resulting friction coefficient tensor for the C240 fullerene model with
purely repulsive and attractive solute-solvent interactions is plotted
in Fig. \ref{fig:tensor}.  It can be seen that the friction coefficient is primarily
perturbed in the direction of the relative coordinate, and the
perpendicular components of the self sub-matrix ($\zeta_{11,\|}$) have
limited spatial dependence and their values are closer to the single
body result.  Concurrently, the value of $\zeta_{12,\bot}$ remains
near zero.

There is less than a 10\% error in the estimation of the friction coefficient
provided by Ref. \onlinecite{Bocquet:1997p103}.  Estimates of error have
been obtained from the variance of the measured plateau or from the
difference between elements of the full friction coefficient tensor that are
equivalent by symmetry.

%


\begin{thebibliography}{71}%
\makeatletter
\providecommand \@ifxundefined [1]{%
 \@ifx{#1\undefined}
}%
\providecommand \@ifnum [1]{%
 \ifnum #1\expandafter \@firstoftwo
 \else \expandafter \@secondoftwo
 \fi
}%
\providecommand \@ifx [1]{%
 \ifx #1\expandafter \@firstoftwo
 \else \expandafter \@secondoftwo
 \fi
}%
\providecommand \natexlab [1]{#1}%
\providecommand \enquote  [1]{``#1''}%
\providecommand \bibnamefont  [1]{#1}%
\providecommand \bibfnamefont [1]{#1}%
\providecommand \citenamefont [1]{#1}%
\providecommand \href@noop [0]{\@secondoftwo}%
\providecommand \href [0]{\begingroup \@sanitize@url \@href}%
\providecommand \@href[1]{\@@startlink{#1}\@@href}%
\providecommand \@@href[1]{\endgroup#1\@@endlink}%
\providecommand \@sanitize@url [0]{\catcode `\\12\catcode `\$12\catcode
  `\&12\catcode `\#12\catcode `\^12\catcode `\_12\catcode `\%12\relax}%
\providecommand \@@startlink[1]{}%
\providecommand \@@endlink[0]{}%
\providecommand \url  [0]{\begingroup\@sanitize@url \@url }%
\providecommand \@url [1]{\endgroup\@href {#1}{\urlprefix }}%
\providecommand \urlprefix  [0]{URL }%
\providecommand \Eprint [0]{\href }%
\providecommand \doibase [0]{http://dx.doi.org/}%
\providecommand \selectlanguage [0]{\@gobble}%
\providecommand \bibinfo  [0]{\@secondoftwo}%
\providecommand \bibfield  [0]{\@secondoftwo}%
\providecommand \translation [1]{[#1]}%
\providecommand \BibitemOpen [0]{}%
\providecommand \bibitemStop [0]{}%
\providecommand \bibitemNoStop [0]{.\EOS\space}%
\providecommand \EOS [0]{\spacefactor3000\relax}%
\providecommand \BibitemShut  [1]{\csname bibitem#1\endcsname}%
\let\auto@bib@innerbib\@empty
\bibitem [{\citenamefont {Pangali}\ \emph {et~al.}(1979)\citenamefont
  {Pangali}, \citenamefont {Rao},\ and\ \citenamefont {Berne}}]{pangali79}%
  \BibitemOpen
  \bibfield  {author} {\bibinfo {author} {\bibfnamefont {C.}~\bibnamefont
  {Pangali}}, \bibinfo {author} {\bibfnamefont {M.}~\bibnamefont {Rao}}, \ and\
  \bibinfo {author} {\bibfnamefont {B.~J.}\ \bibnamefont {Berne}},\ }\href@noop
  {} {\bibfield  {journal} {\bibinfo  {journal} {J. Chem. Phys.}\ }\textbf
  {\bibinfo {volume} {71}},\ \bibinfo {pages} {2982} (\bibinfo {year}
  {1979})}\BibitemShut {NoStop}%
\bibitem [{\citenamefont {Chandler}(2005)}]{Chandler:2005p75}%
  \BibitemOpen
  \bibfield  {author} {\bibinfo {author} {\bibfnamefont {D.}~\bibnamefont
  {Chandler}},\ }\href@noop {} {\bibfield  {journal} {\bibinfo  {journal}
  {Nature}\ }\textbf {\bibinfo {volume} {437}},\ \bibinfo {pages} {640}
  (\bibinfo {year} {2005})}\BibitemShut {NoStop}%
\bibitem [{\citenamefont {Berne}\ \emph {et~al.}(2009)\citenamefont {Berne},
  \citenamefont {Weeks},\ and\ \citenamefont {Zhou}}]{Berne:2009dg}%
  \BibitemOpen
  \bibfield  {author} {\bibinfo {author} {\bibfnamefont {B.~J.}\ \bibnamefont
  {Berne}}, \bibinfo {author} {\bibfnamefont {J.~D.}\ \bibnamefont {Weeks}}, \
  and\ \bibinfo {author} {\bibfnamefont {R.}~\bibnamefont {Zhou}},\ }\href@noop
  {} {\bibfield  {journal} {\bibinfo  {journal} {Annu. Rev. Phys. Chem.}\
  }\textbf {\bibinfo {volume} {60}},\ \bibinfo {pages} {85} (\bibinfo {year}
  {2009})}\BibitemShut {NoStop}%
\bibitem [{\citenamefont {Jamadagni}\ \emph {et~al.}(2011)\citenamefont
  {Jamadagni}, \citenamefont {Godawat},\ and\ \citenamefont
  {Garde}}]{Jamadagni:2010bc}%
  \BibitemOpen
  \bibfield  {author} {\bibinfo {author} {\bibfnamefont {S.~N.}\ \bibnamefont
  {Jamadagni}}, \bibinfo {author} {\bibfnamefont {R.}~\bibnamefont {Godawat}},
  \ and\ \bibinfo {author} {\bibfnamefont {S.}~\bibnamefont {Garde}},\
  }\href@noop {} {\bibfield  {journal} {\bibinfo  {journal} {Annu. Rev. Chem.
  and Biomol. Eng.}\ }\textbf {\bibinfo {volume} {2}},\ \bibinfo {pages} {147}
  (\bibinfo {year} {2011})}\BibitemShut {NoStop}%
\bibitem [{\citenamefont {Wallqvist}\ and\ \citenamefont
  {Berne}(1995)}]{WALLQVIST:1995uw}%
  \BibitemOpen
  \bibfield  {author} {\bibinfo {author} {\bibfnamefont {A.}~\bibnamefont
  {Wallqvist}}\ and\ \bibinfo {author} {\bibfnamefont {B.~J.}\ \bibnamefont
  {Berne}},\ }\href@noop {} {\bibfield  {journal} {\bibinfo  {journal} {J.
  Phys. Chem.}\ }\textbf {\bibinfo {volume} {99}},\ \bibinfo {pages} {2893}
  (\bibinfo {year} {1995})}\BibitemShut {NoStop}%
\bibitem [{\citenamefont {Hummer}\ \emph {et~al.}(1996)\citenamefont {Hummer},
  \citenamefont {Garde}, \citenamefont {Garc{\'\i}a}, \citenamefont
  {Pohorille},\ and\ \citenamefont {Pratt}}]{hummer96}%
  \BibitemOpen
  \bibfield  {author} {\bibinfo {author} {\bibfnamefont {G.}~\bibnamefont
  {Hummer}}, \bibinfo {author} {\bibfnamefont {S.}~\bibnamefont {Garde}},
  \bibinfo {author} {\bibfnamefont {A.~E.}\ \bibnamefont {Garc{\'\i}a}},
  \bibinfo {author} {\bibfnamefont {A.}~\bibnamefont {Pohorille}}, \ and\
  \bibinfo {author} {\bibfnamefont {L.~R.}\ \bibnamefont {Pratt}},\ }\href@noop
  {} {\bibfield  {journal} {\bibinfo  {journal} {Proc. Natl. Acad. Sci.}\
  }\textbf {\bibinfo {volume} {93}},\ \bibinfo {pages} {8951} (\bibinfo {year}
  {1996})}\BibitemShut {NoStop}%
\bibitem [{\citenamefont {Lum}\ \emph {et~al.}(1999)\citenamefont {Lum},
  \citenamefont {Chandler},\ and\ \citenamefont {Weeks}}]{Lum:1999p624}%
  \BibitemOpen
  \bibfield  {author} {\bibinfo {author} {\bibfnamefont {K.}~\bibnamefont
  {Lum}}, \bibinfo {author} {\bibfnamefont {D.}~\bibnamefont {Chandler}}, \
  and\ \bibinfo {author} {\bibfnamefont {J.}~\bibnamefont {Weeks}},\
  }\href@noop {} {\bibfield  {journal} {\bibinfo  {journal} {J. Phys. Chem. B}\
  }\textbf {\bibinfo {volume} {103}},\ \bibinfo {pages} {4570} (\bibinfo {year}
  {1999})}\BibitemShut {NoStop}%
\bibitem [{\citenamefont {Huang}\ and\ \citenamefont
  {Chandler}(2002)}]{huang02}%
  \BibitemOpen
  \bibfield  {author} {\bibinfo {author} {\bibfnamefont {D.~M.}\ \bibnamefont
  {Huang}}\ and\ \bibinfo {author} {\bibfnamefont {D.}~\bibnamefont
  {Chandler}},\ }\href@noop {} {\bibfield  {journal} {\bibinfo  {journal} {J.
  Phys. Chem. B}\ }\textbf {\bibinfo {volume} {106}},\ \bibinfo {pages} {2047}
  (\bibinfo {year} {2002})}\BibitemShut {NoStop}%
\bibitem [{\citenamefont {Ten~Wolde}\ and\ \citenamefont
  {Chandler}(2002)}]{TenWolde:2002p71}%
  \BibitemOpen
  \bibfield  {author} {\bibinfo {author} {\bibfnamefont {P.~R.}\ \bibnamefont
  {Ten~Wolde}}\ and\ \bibinfo {author} {\bibfnamefont {D.}~\bibnamefont
  {Chandler}},\ }\href@noop {} {\bibfield  {journal} {\bibinfo  {journal}
  {Proc. Natl. Acad. Sci.}\ }\textbf {\bibinfo {volume} {99}},\ \bibinfo
  {pages} {6539} (\bibinfo {year} {2002})}\BibitemShut {NoStop}%
\bibitem [{\citenamefont {Huang}\ \emph {et~al.}(2003)\citenamefont {Huang},
  \citenamefont {Margulis},\ and\ \citenamefont {Berne}}]{Margulis:2003wo}%
  \BibitemOpen
  \bibfield  {author} {\bibinfo {author} {\bibfnamefont {X.}~\bibnamefont
  {Huang}}, \bibinfo {author} {\bibfnamefont {C.~J.}\ \bibnamefont {Margulis}},
  \ and\ \bibinfo {author} {\bibfnamefont {B.~J.}\ \bibnamefont {Berne}},\
  }\href@noop {} {\bibfield  {journal} {\bibinfo  {journal} {Proc. Natl. Acad.
  Sci.}\ }\textbf {\bibinfo {volume} {100}},\ \bibinfo {pages} {11953}
  (\bibinfo {year} {2003})}\BibitemShut {NoStop}%
\bibitem [{\citenamefont {Zhou}\ \emph {et~al.}(2004)\citenamefont {Zhou},
  \citenamefont {Huang}, \citenamefont {Margulis},\ and\ \citenamefont
  {Berne}}]{ruhong04}%
  \BibitemOpen
  \bibfield  {author} {\bibinfo {author} {\bibfnamefont {R.}~\bibnamefont
  {Zhou}}, \bibinfo {author} {\bibfnamefont {X.}~\bibnamefont {Huang}},
  \bibinfo {author} {\bibfnamefont {C.~J.}\ \bibnamefont {Margulis}}, \ and\
  \bibinfo {author} {\bibfnamefont {B.~J.}\ \bibnamefont {Berne}},\ }\href@noop
  {} {\bibfield  {journal} {\bibinfo  {journal} {Science}\ }\textbf {\bibinfo
  {volume} {305}},\ \bibinfo {pages} {1605} (\bibinfo {year}
  {2004})}\BibitemShut {NoStop}%
\bibitem [{\citenamefont {Huang}\ \emph {et~al.}(2005)\citenamefont {Huang},
  \citenamefont {Zhou},\ and\ \citenamefont {Berne}}]{Huang:2005ho}%
  \BibitemOpen
  \bibfield  {author} {\bibinfo {author} {\bibfnamefont {X.}~\bibnamefont
  {Huang}}, \bibinfo {author} {\bibfnamefont {R.}~\bibnamefont {Zhou}}, \ and\
  \bibinfo {author} {\bibfnamefont {B.~J.}\ \bibnamefont {Berne}},\ }\href@noop
  {} {\bibfield  {journal} {\bibinfo  {journal} {J. Phys. Chem. B}\ }\textbf
  {\bibinfo {volume} {109}},\ \bibinfo {pages} {3546} (\bibinfo {year}
  {2005})}\BibitemShut {NoStop}%
\bibitem [{\citenamefont {Giovambattista}\ \emph {et~al.}(2007)\citenamefont
  {Giovambattista}, \citenamefont {Debenedetti},\ and\ \citenamefont
  {Rossky}}]{nicolas07}%
  \BibitemOpen
  \bibfield  {author} {\bibinfo {author} {\bibfnamefont {N.}~\bibnamefont
  {Giovambattista}}, \bibinfo {author} {\bibfnamefont {P.~G.}\ \bibnamefont
  {Debenedetti}}, \ and\ \bibinfo {author} {\bibfnamefont {P.~J.}\ \bibnamefont
  {Rossky}},\ }\href@noop {} {\bibfield  {journal} {\bibinfo  {journal} {J.
  Phys. Chem. B}\ }\textbf {\bibinfo {volume} {111}},\ \bibinfo {pages} {9581}
  (\bibinfo {year} {2007})}\BibitemShut {NoStop}%
\bibitem [{\citenamefont {Mittal}\ and\ \citenamefont
  {Hummer}(2008)}]{mittal2008}%
  \BibitemOpen
  \bibfield  {author} {\bibinfo {author} {\bibfnamefont {J.}~\bibnamefont
  {Mittal}}\ and\ \bibinfo {author} {\bibfnamefont {G.}~\bibnamefont
  {Hummer}},\ }\href@noop {} {\bibfield  {journal} {\bibinfo  {journal} {Proc.
  Natl. Acad. Sci.}\ }\textbf {\bibinfo {volume} {105}},\ \bibinfo {pages}
  {20130} (\bibinfo {year} {2008})}\BibitemShut {NoStop}%
\bibitem [{\citenamefont {Willard}\ and\ \citenamefont
  {Chandler}(2008)}]{Willard:2008p84}%
  \BibitemOpen
  \bibfield  {author} {\bibinfo {author} {\bibfnamefont {A.~P.}\ \bibnamefont
  {Willard}}\ and\ \bibinfo {author} {\bibfnamefont {D.}~\bibnamefont
  {Chandler}},\ }\href@noop {} {\bibfield  {journal} {\bibinfo  {journal} {J.
  Phys. Chem. B}\ }\textbf {\bibinfo {volume} {112}},\ \bibinfo {pages} {6187}
  (\bibinfo {year} {2008})}\BibitemShut {NoStop}%
\bibitem [{\citenamefont {Sarupria}\ and\ \citenamefont
  {Garde}(2009)}]{Sarupria:2009dq}%
  \BibitemOpen
  \bibfield  {author} {\bibinfo {author} {\bibfnamefont {S.}~\bibnamefont
  {Sarupria}}\ and\ \bibinfo {author} {\bibfnamefont {S.}~\bibnamefont
  {Garde}},\ }\href@noop {} {\bibfield  {journal} {\bibinfo  {journal} {Phys.
  Rev. Lett.}\ }\textbf {\bibinfo {volume} {103}},\ \bibinfo {pages} {037803}
  (\bibinfo {year} {2009})}\BibitemShut {NoStop}%
\bibitem [{\citenamefont {Godawat}\ \emph {et~al.}(2009)\citenamefont
  {Godawat}, \citenamefont {Jamadagni},\ and\ \citenamefont
  {Garde}}]{godawat09}%
  \BibitemOpen
  \bibfield  {author} {\bibinfo {author} {\bibfnamefont {R.}~\bibnamefont
  {Godawat}}, \bibinfo {author} {\bibfnamefont {S.~N.}\ \bibnamefont
  {Jamadagni}}, \ and\ \bibinfo {author} {\bibfnamefont {S.}~\bibnamefont
  {Garde}},\ }\href@noop {} {\bibfield  {journal} {\bibinfo  {journal} {Proc.
  Natl. Acad. Sci.}\ }\textbf {\bibinfo {volume} {106}},\ \bibinfo {pages}
  {15119} (\bibinfo {year} {2009})}\BibitemShut {NoStop}%
\bibitem [{\citenamefont {Patel}\ \emph {et~al.}(2010)\citenamefont {Patel},
  \citenamefont {Varilly},\ and\ \citenamefont {Chandler}}]{patel10}%
  \BibitemOpen
  \bibfield  {author} {\bibinfo {author} {\bibfnamefont {A.~J.}\ \bibnamefont
  {Patel}}, \bibinfo {author} {\bibfnamefont {P.}~\bibnamefont {Varilly}}, \
  and\ \bibinfo {author} {\bibfnamefont {D.}~\bibnamefont {Chandler}},\
  }\href@noop {} {\bibfield  {journal} {\bibinfo  {journal} {J. Phys. Chem. B}\
  }\textbf {\bibinfo {volume} {114}},\ \bibinfo {pages} {1632} (\bibinfo {year}
  {2010})}\BibitemShut {NoStop}%
\bibitem [{\citenamefont {Huang}\ \emph {et~al.}(2008)\citenamefont {Huang},
  \citenamefont {Sendner}, \citenamefont {Horinek}, \citenamefont {Netz},\ and\
  \citenamefont {Bocquet}}]{Sendner:2008tg}%
  \BibitemOpen
  \bibfield  {author} {\bibinfo {author} {\bibfnamefont {D.~M.}\ \bibnamefont
  {Huang}}, \bibinfo {author} {\bibfnamefont {C.}~\bibnamefont {Sendner}},
  \bibinfo {author} {\bibfnamefont {D.}~\bibnamefont {Horinek}}, \bibinfo
  {author} {\bibfnamefont {R.~R.}\ \bibnamefont {Netz}}, \ and\ \bibinfo
  {author} {\bibfnamefont {L.}~\bibnamefont {Bocquet}},\ }\href@noop {}
  {\bibfield  {journal} {\bibinfo  {journal} {Phys. Rev. Lett.}\ }\textbf
  {\bibinfo {volume} {101}},\ \bibinfo {pages} {226101} (\bibinfo {year}
  {2008})}\BibitemShut {NoStop}%
\bibitem [{\citenamefont {Sendner}\ \emph {et~al.}(2009)\citenamefont
  {Sendner}, \citenamefont {Horinek}, \citenamefont {Bocquet},\ and\
  \citenamefont {Netz}}]{Sendner:2009gi}%
  \BibitemOpen
  \bibfield  {author} {\bibinfo {author} {\bibfnamefont {C.}~\bibnamefont
  {Sendner}}, \bibinfo {author} {\bibfnamefont {D.}~\bibnamefont {Horinek}},
  \bibinfo {author} {\bibfnamefont {L.}~\bibnamefont {Bocquet}}, \ and\
  \bibinfo {author} {\bibfnamefont {R.~R.}\ \bibnamefont {Netz}},\ }\href@noop
  {} {\bibfield  {journal} {\bibinfo  {journal} {Langmuir}\ }\textbf {\bibinfo
  {volume} {25}},\ \bibinfo {pages} {10768} (\bibinfo {year}
  {2009})}\BibitemShut {NoStop}%
\bibitem [{\citenamefont {Thomas}\ and\ \citenamefont
  {McGaughey}(2009)}]{Thomas:2009tj}%
  \BibitemOpen
  \bibfield  {author} {\bibinfo {author} {\bibfnamefont {J.~A.}\ \bibnamefont
  {Thomas}}\ and\ \bibinfo {author} {\bibfnamefont {A.~J.~H.}\ \bibnamefont
  {McGaughey}},\ }\href@noop {} {\bibfield  {journal} {\bibinfo  {journal}
  {Phys. Rev. Lett.}\ }\textbf {\bibinfo {volume} {102}},\ \bibinfo {pages}
  {184502} (\bibinfo {year} {2009})}\BibitemShut {NoStop}%
\bibitem [{\citenamefont {Kalra}\ \emph {et~al.}(2010)\citenamefont {Kalra},
  \citenamefont {Garde},\ and\ \citenamefont {Hummer}}]{Kalra:2010p231}%
  \BibitemOpen
  \bibfield  {author} {\bibinfo {author} {\bibfnamefont {A.}~\bibnamefont
  {Kalra}}, \bibinfo {author} {\bibfnamefont {S.}~\bibnamefont {Garde}}, \ and\
  \bibinfo {author} {\bibfnamefont {G.}~\bibnamefont {Hummer}},\ }\href@noop {}
  {\bibfield  {journal} {\bibinfo  {journal} {Euro. Phys. J. Special Topics}\
  }\textbf {\bibinfo {volume} {189}},\ \bibinfo {pages} {147} (\bibinfo {year}
  {2010})}\BibitemShut {NoStop}%
\bibitem [{\citenamefont {Falk}\ \emph {et~al.}(2010)\citenamefont {Falk},
  \citenamefont {Sedlmeier}, \citenamefont {Joly}, \citenamefont {Netz},\ and\
  \citenamefont {Bocquet}}]{Falk:2010cr}%
  \BibitemOpen
  \bibfield  {author} {\bibinfo {author} {\bibfnamefont {K.}~\bibnamefont
  {Falk}}, \bibinfo {author} {\bibfnamefont {F.}~\bibnamefont {Sedlmeier}},
  \bibinfo {author} {\bibfnamefont {L.}~\bibnamefont {Joly}}, \bibinfo {author}
  {\bibfnamefont {R.~R.}\ \bibnamefont {Netz}}, \ and\ \bibinfo {author}
  {\bibfnamefont {L.}~\bibnamefont {Bocquet}},\ }\href@noop {} {\bibfield
  {journal} {\bibinfo  {journal} {Nano Lett.}\ }\textbf {\bibinfo {volume}
  {10}},\ \bibinfo {pages} {4067} (\bibinfo {year} {2010})}\BibitemShut
  {NoStop}%
\bibitem [{\citenamefont {Bocquet}\ and\ \citenamefont
  {Charlaix}(2010)}]{Bocquet:cy}%
  \BibitemOpen
  \bibfield  {author} {\bibinfo {author} {\bibfnamefont {L.}~\bibnamefont
  {Bocquet}}\ and\ \bibinfo {author} {\bibfnamefont {E.}~\bibnamefont
  {Charlaix}},\ }\href@noop {} {\bibfield  {journal} {\bibinfo  {journal}
  {Chem. Soc. Rev.}\ }\textbf {\bibinfo {volume} {39}},\ \bibinfo {pages}
  {1073} (\bibinfo {year} {2010})}\BibitemShut {NoStop}%
\bibitem [{\citenamefont {Ermak}\ and\ \citenamefont
  {McCammon}(1978)}]{mccammon78}%
  \BibitemOpen
  \bibfield  {author} {\bibinfo {author} {\bibfnamefont {D.~L.}\ \bibnamefont
  {Ermak}}\ and\ \bibinfo {author} {\bibfnamefont {J.~A.}\ \bibnamefont
  {McCammon}},\ }\href@noop {} {\bibfield  {journal} {\bibinfo  {journal} {J.
  Chem. Phys.}\ }\textbf {\bibinfo {volume} {69}},\ \bibinfo {pages} {1352}
  (\bibinfo {year} {1978})}\BibitemShut {NoStop}%
\bibitem [{\citenamefont {Brady}\ \emph {et~al.}(1988)\citenamefont {Brady},
  \citenamefont {Philips}, \citenamefont {Lester},\ and\ \citenamefont
  {Bossis}}]{Brady:1988tz}%
  \BibitemOpen
  \bibfield  {author} {\bibinfo {author} {\bibfnamefont {J.~F.}\ \bibnamefont
  {Brady}}, \bibinfo {author} {\bibfnamefont {R.~J.}\ \bibnamefont {Philips}},
  \bibinfo {author} {\bibfnamefont {J.~C.}\ \bibnamefont {Lester}}, \ and\
  \bibinfo {author} {\bibfnamefont {G.}~\bibnamefont {Bossis}},\ }\href@noop {}
  {\bibfield  {journal} {\bibinfo  {journal} {J. Fluid Mech.}\ }\textbf
  {\bibinfo {volume} {195}},\ \bibinfo {pages} {257} (\bibinfo {year}
  {1988})}\BibitemShut {NoStop}%
\bibitem [{\citenamefont {Brady}(1988)}]{Brady:1988p970}%
  \BibitemOpen
  \bibfield  {author} {\bibinfo {author} {\bibfnamefont {J.}~\bibnamefont
  {Brady}},\ }\href@noop {} {\bibfield  {journal} {\bibinfo  {journal} {Annu.
  Rev. Fluid Mech}\ }\textbf {\bibinfo {volume} {20}},\ \bibinfo {pages} {111}
  (\bibinfo {year} {1988})}\BibitemShut {NoStop}%
\bibitem [{\citenamefont {Chatterji}\ and\ \citenamefont
  {Horbach}(2005)}]{chatterji05}%
  \BibitemOpen
  \bibfield  {author} {\bibinfo {author} {\bibfnamefont {A.}~\bibnamefont
  {Chatterji}}\ and\ \bibinfo {author} {\bibfnamefont {J.}~\bibnamefont
  {Horbach}},\ }\href@noop {} {\bibfield  {journal} {\bibinfo  {journal} {J.
  Chem. Phys.}\ }\textbf {\bibinfo {volume} {122}},\ \bibinfo {pages} {184903}
  (\bibinfo {year} {2005})}\BibitemShut {NoStop}%
\bibitem [{\citenamefont {Padding}\ and\ \citenamefont
  {Louis}(2006)}]{padding06}%
  \BibitemOpen
  \bibfield  {author} {\bibinfo {author} {\bibfnamefont {J.~T.}\ \bibnamefont
  {Padding}}\ and\ \bibinfo {author} {\bibfnamefont {A.~A.}\ \bibnamefont
  {Louis}},\ }\href@noop {} {\bibfield  {journal} {\bibinfo  {journal} {Phys.
  Rev. E}\ }\textbf {\bibinfo {volume} {74}},\ \bibinfo {pages} {031402}
  (\bibinfo {year} {2006})}\BibitemShut {NoStop}%
\bibitem [{\citenamefont {Ando}\ and\ \citenamefont
  {Skolnick}(2010)}]{Ando:2010p99}%
  \BibitemOpen
  \bibfield  {author} {\bibinfo {author} {\bibfnamefont {T.}~\bibnamefont
  {Ando}}\ and\ \bibinfo {author} {\bibfnamefont {J.}~\bibnamefont
  {Skolnick}},\ }\href@noop {} {\bibfield  {journal} {\bibinfo  {journal}
  {Proc. Natl. Acad. Sci.}\ }\textbf {\bibinfo {volume} {107}},\ \bibinfo
  {pages} {18457} (\bibinfo {year} {2010})}\BibitemShut {NoStop}%
\bibitem [{\citenamefont {Berne}\ and\ \citenamefont
  {Harp}(1970)}]{Berne:1970p90}%
  \BibitemOpen
  \bibfield  {author} {\bibinfo {author} {\bibfnamefont {B.~J.}\ \bibnamefont
  {Berne}}\ and\ \bibinfo {author} {\bibfnamefont {G.~D.}\ \bibnamefont
  {Harp}},\ }\href@noop {} {\bibfield  {journal} {\bibinfo  {journal} {Adv.
  Chem . Phys.}\ }\textbf {\bibinfo {volume} {17}},\ \bibinfo {pages} {1}
  (\bibinfo {year} {1970})}\BibitemShut {NoStop}%
\bibitem [{\citenamefont {Straub}\ \emph {et~al.}(1987)\citenamefont {Straub},
  \citenamefont {Borkovec},\ and\ \citenamefont {Berne}}]{Straub:1987p76}%
  \BibitemOpen
  \bibfield  {author} {\bibinfo {author} {\bibfnamefont {J.~E.}\ \bibnamefont
  {Straub}}, \bibinfo {author} {\bibfnamefont {M.}~\bibnamefont {Borkovec}}, \
  and\ \bibinfo {author} {\bibfnamefont {B.~J.}\ \bibnamefont {Berne}},\
  }\href@noop {} {\bibfield  {journal} {\bibinfo  {journal} {J. Phys. Chem.}\
  }\textbf {\bibinfo {volume} {91}},\ \bibinfo {pages} {4995} (\bibinfo {year}
  {1987})}\BibitemShut {NoStop}%
\bibitem [{\citenamefont {Straub}\ \emph {et~al.}(1990)\citenamefont {Straub},
  \citenamefont {Berne},\ and\ \citenamefont {Roux}}]{Straub:1990p65}%
  \BibitemOpen
  \bibfield  {author} {\bibinfo {author} {\bibfnamefont {J.~E.}\ \bibnamefont
  {Straub}}, \bibinfo {author} {\bibfnamefont {B.~J.}\ \bibnamefont {Berne}}, \
  and\ \bibinfo {author} {\bibfnamefont {B.}~\bibnamefont {Roux}},\ }\href@noop
  {} {\bibfield  {journal} {\bibinfo  {journal} {J. Chem. Phys.}\ }\textbf
  {\bibinfo {volume} {93}},\ \bibinfo {pages} {6804} (\bibinfo {year}
  {1990})}\BibitemShut {NoStop}%
\bibitem [{\citenamefont {Espa{\~n}ol}\ and\ \citenamefont
  {Z{\'u}{\~n}iga}(1993)}]{Espanol:1993p63}%
  \BibitemOpen
  \bibfield  {author} {\bibinfo {author} {\bibfnamefont {P.}~\bibnamefont
  {Espa{\~n}ol}}\ and\ \bibinfo {author} {\bibfnamefont {I.}~\bibnamefont
  {Z{\'u}{\~n}iga}},\ }\href@noop {} {\bibfield  {journal} {\bibinfo  {journal}
  {J. Chem. Phys.}\ }\textbf {\bibinfo {volume} {98}},\ \bibinfo {pages} {574}
  (\bibinfo {year} {1993})}\BibitemShut {NoStop}%
\bibitem [{\citenamefont {Bocquet}\ \emph {et~al.}(1994)\citenamefont
  {Bocquet}, \citenamefont {Hansen},\ and\ \citenamefont
  {Piasecki}}]{Bocquet:1994p317}%
  \BibitemOpen
  \bibfield  {author} {\bibinfo {author} {\bibfnamefont {L.}~\bibnamefont
  {Bocquet}}, \bibinfo {author} {\bibfnamefont {J.}~\bibnamefont {Hansen}}, \
  and\ \bibinfo {author} {\bibfnamefont {J.}~\bibnamefont {Piasecki}},\
  }\href@noop {} {\bibfield  {journal} {\bibinfo  {journal} {J. Stat. Phys.}\
  }\textbf {\bibinfo {volume} {76}},\ \bibinfo {pages} {527} (\bibinfo {year}
  {1994})}\BibitemShut {NoStop}%
\bibitem [{\citenamefont {Bocquet}\ \emph {et~al.}(1997)\citenamefont
  {Bocquet}, \citenamefont {Hansen},\ and\ \citenamefont
  {Piasecki}}]{Bocquet:1997p103}%
  \BibitemOpen
  \bibfield  {author} {\bibinfo {author} {\bibfnamefont {L.}~\bibnamefont
  {Bocquet}}, \bibinfo {author} {\bibfnamefont {J.}~\bibnamefont {Hansen}}, \
  and\ \bibinfo {author} {\bibfnamefont {J.}~\bibnamefont {Piasecki}},\
  }\href@noop {} {\bibfield  {journal} {\bibinfo  {journal} {J. Stat. Phys.}\
  }\textbf {\bibinfo {volume} {89}},\ \bibinfo {pages} {321} (\bibinfo {year}
  {1997})}\BibitemShut {NoStop}%
\bibitem [{\citenamefont {Lee}(2010)}]{Lee:2010p307}%
  \BibitemOpen
  \bibfield  {author} {\bibinfo {author} {\bibfnamefont {S.~H.}\ \bibnamefont
  {Lee}},\ }\href@noop {} {\bibfield  {journal} {\bibinfo  {journal} {Bull.
  Korean Chem. Soc.}\ }\textbf {\bibinfo {volume} {31}},\ \bibinfo {pages}
  {2402} (\bibinfo {year} {2010})}\BibitemShut {NoStop}%
\bibitem [{\citenamefont {Bezmel'nitsyn}\ \emph {et~al.}(1998)\citenamefont
  {Bezmel'nitsyn}, \citenamefont {Eletskii},\ and\ \citenamefont
  {Okun}}]{fulsol}%
  \BibitemOpen
  \bibfield  {author} {\bibinfo {author} {\bibfnamefont {V.~N.}\ \bibnamefont
  {Bezmel'nitsyn}}, \bibinfo {author} {\bibfnamefont {A.~V.}\ \bibnamefont
  {Eletskii}}, \ and\ \bibinfo {author} {\bibfnamefont {M.~V.}\ \bibnamefont
  {Okun}},\ }\href@noop {} {\bibfield  {journal} {\bibinfo  {journal} {Physics
  - Uspekhi}\ }\textbf {\bibinfo {volume} {41}},\ \bibinfo {pages} {1091}
  (\bibinfo {year} {1998})}\BibitemShut {NoStop}%
\bibitem [{\citenamefont {Hotta}\ \emph {et~al.}(2005)\citenamefont {Hotta},
  \citenamefont {Kimura},\ and\ \citenamefont {Sasai}}]{Hotta:2005bp}%
  \BibitemOpen
  \bibfield  {author} {\bibinfo {author} {\bibfnamefont {T.}~\bibnamefont
  {Hotta}}, \bibinfo {author} {\bibfnamefont {A.}~\bibnamefont {Kimura}}, \
  and\ \bibinfo {author} {\bibfnamefont {M.}~\bibnamefont {Sasai}},\
  }\href@noop {} {\bibfield  {journal} {\bibinfo  {journal} {J. Phys. Chem. B}\
  }\textbf {\bibinfo {volume} {109}},\ \bibinfo {pages} {18600} (\bibinfo
  {year} {2005})}\BibitemShut {NoStop}%
\bibitem [{\citenamefont {Makowski}\ \emph {et~al.}(2009)\citenamefont
  {Makowski}, \citenamefont {Czaplewski}, \citenamefont {Liwo},\ and\
  \citenamefont {Scheraga}}]{Makowski:2009p61}%
  \BibitemOpen
  \bibfield  {author} {\bibinfo {author} {\bibfnamefont {M.}~\bibnamefont
  {Makowski}}, \bibinfo {author} {\bibfnamefont {C.}~\bibnamefont
  {Czaplewski}}, \bibinfo {author} {\bibfnamefont {A.}~\bibnamefont {Liwo}}, \
  and\ \bibinfo {author} {\bibfnamefont {H.~A.}\ \bibnamefont {Scheraga}},\
  }\href@noop {} {\bibfield  {journal} {\bibinfo  {journal} {J. Phys. Chem. B}\
  }\textbf {\bibinfo {volume} {114}},\ \bibinfo {pages} {993} (\bibinfo {year}
  {2009})}\BibitemShut {NoStop}%
\bibitem [{\citenamefont {Emeis}\ and\ \citenamefont {Fehder}(1970)}]{emeis70}%
  \BibitemOpen
  \bibfield  {author} {\bibinfo {author} {\bibfnamefont {C.~A.}\ \bibnamefont
  {Emeis}}\ and\ \bibinfo {author} {\bibfnamefont {P.~L.}\ \bibnamefont
  {Fehder}},\ }\href@noop {} {\bibfield  {journal} {\bibinfo  {journal} {J. Am.
  Chem. Soc.}\ }\textbf {\bibinfo {volume} {92}},\ \bibinfo {pages} {2246}
  (\bibinfo {year} {1970})}\BibitemShut {NoStop}%
\bibitem [{\citenamefont {Deutch}\ and\ \citenamefont
  {Felderhof}(1973)}]{deutch73}%
  \BibitemOpen
  \bibfield  {author} {\bibinfo {author} {\bibfnamefont {J.~M.}\ \bibnamefont
  {Deutch}}\ and\ \bibinfo {author} {\bibfnamefont {B.~U.}\ \bibnamefont
  {Felderhof}},\ }\href@noop {} {\bibfield  {journal} {\bibinfo  {journal} {J.
  Chem. Phys.}\ }\textbf {\bibinfo {volume} {59}},\ \bibinfo {pages} {1669}
  (\bibinfo {year} {1973})}\BibitemShut {NoStop}%
\bibitem [{\citenamefont {Wolynes}\ and\ \citenamefont
  {Deutch}(1976)}]{wolynes76}%
  \BibitemOpen
  \bibfield  {author} {\bibinfo {author} {\bibfnamefont {P.~G.}\ \bibnamefont
  {Wolynes}}\ and\ \bibinfo {author} {\bibfnamefont {J.~M.}\ \bibnamefont
  {Deutch}},\ }\href@noop {} {\bibfield  {journal} {\bibinfo  {journal} {J.
  Chem. Phys.}\ }\textbf {\bibinfo {volume} {65}},\ \bibinfo {pages} {450}
  (\bibinfo {year} {1976})}\BibitemShut {NoStop}%
\bibitem [{\citenamefont {Northrup}\ and\ \citenamefont
  {Hynes}(1979)}]{Northrup:1979wr}%
  \BibitemOpen
  \bibfield  {author} {\bibinfo {author} {\bibfnamefont {S.~H.}\ \bibnamefont
  {Northrup}}\ and\ \bibinfo {author} {\bibfnamefont {J.~T.}\ \bibnamefont
  {Hynes}},\ }\href@noop {} {\bibfield  {journal} {\bibinfo  {journal} {J.
  Chem. Phys.}\ }\textbf {\bibinfo {volume} {71}},\ \bibinfo {pages} {871}
  (\bibinfo {year} {1979})}\BibitemShut {NoStop}%
\bibitem [{\citenamefont {Calef}\ and\ \citenamefont {Deutch}(1983)}]{calef83}%
  \BibitemOpen
  \bibfield  {author} {\bibinfo {author} {\bibfnamefont {D.~F.}\ \bibnamefont
  {Calef}}\ and\ \bibinfo {author} {\bibfnamefont {J.~M.}\ \bibnamefont
  {Deutch}},\ }\href@noop {} {\bibfield  {journal} {\bibinfo  {journal} {Annu.
  Rev. Phys. Chem.}\ }\textbf {\bibinfo {volume} {34}},\ \bibinfo {pages} {493}
  (\bibinfo {year} {1983})}\BibitemShut {NoStop}%
\bibitem [{\citenamefont {Dhont}(1996)}]{dhont}%
  \BibitemOpen
  \bibfield  {author} {\bibinfo {author} {\bibfnamefont {J.~K.~G.}\
  \bibnamefont {Dhont}},\ }\href@noop {} {\emph {\bibinfo {title} {An
  Introduction to Dynamics of Colloids}}},\ edited by\ \bibinfo {editor}
  {\bibfnamefont {D.}~\bibnamefont {Mobius}}\ and\ \bibinfo {editor}
  {\bibfnamefont {R.}~\bibnamefont {Miller}},\ \bibinfo {series} {Studies in
  Interface Science}, Vol.~\bibinfo {volume} {2}\ (\bibinfo  {publisher}
  {Elsevier},\ \bibinfo {address} {Amsterdam},\ \bibinfo {year}
  {1996})\BibitemShut {NoStop}%
\bibitem [{\citenamefont {Jeffrey}\ and\ \citenamefont
  {Onishi}(1984)}]{Jeffrey:1984tu}%
  \BibitemOpen
  \bibfield  {author} {\bibinfo {author} {\bibfnamefont {D.~J.}\ \bibnamefont
  {Jeffrey}}\ and\ \bibinfo {author} {\bibfnamefont {Y.}~\bibnamefont
  {Onishi}},\ }\href@noop {} {\bibfield  {journal} {\bibinfo  {journal} {J.
  Fluid Mech.}\ }\textbf {\bibinfo {volume} {139}},\ \bibinfo {pages} {261}
  (\bibinfo {year} {1984})}\BibitemShut {NoStop}%
\bibitem [{\citenamefont {Beenakker}(1986)}]{Beenakker:1986tz}%
  \BibitemOpen
  \bibfield  {author} {\bibinfo {author} {\bibfnamefont {C.}~\bibnamefont
  {Beenakker}},\ }\href@noop {} {\bibfield  {journal} {\bibinfo  {journal} {J.
  Chem. Phys.}\ }\textbf {\bibinfo {volume} {85}},\ \bibinfo {pages} {1581}
  (\bibinfo {year} {1986})}\BibitemShut {NoStop}%
\bibitem [{\citenamefont {Chen}(1998)}]{Chen:1998vv}%
  \BibitemOpen
  \bibfield  {author} {\bibinfo {author} {\bibfnamefont {S.}~\bibnamefont
  {Chen}},\ }\href@noop {} {\bibfield  {journal} {\bibinfo  {journal} {Annu.
  Rev. Fluid Mech.}\ }\textbf {\bibinfo {volume} {30}},\ \bibinfo {pages} {329}
  (\bibinfo {year} {1998})}\BibitemShut {NoStop}%
\bibitem [{\citenamefont {Nakayama}\ \emph {et~al.}(2008)\citenamefont
  {Nakayama}, \citenamefont {Kim},\ and\ \citenamefont
  {Yamamoto}}]{Nakayama:2008p587}%
  \BibitemOpen
  \bibfield  {author} {\bibinfo {author} {\bibfnamefont {Y.}~\bibnamefont
  {Nakayama}}, \bibinfo {author} {\bibfnamefont {K.}~\bibnamefont {Kim}}, \
  and\ \bibinfo {author} {\bibfnamefont {R.}~\bibnamefont {Yamamoto}},\
  }\href@noop {} {\bibfield  {journal} {\bibinfo  {journal} {Euro. Phys. J. E}\
  }\textbf {\bibinfo {volume} {26}},\ \bibinfo {pages} {361} (\bibinfo {year}
  {2008})}\BibitemShut {NoStop}%
\bibitem [{\citenamefont {Voulgarakis}\ and\ \citenamefont
  {Chu}(2009)}]{chu09}%
  \BibitemOpen
  \bibfield  {author} {\bibinfo {author} {\bibfnamefont {N.~K.}\ \bibnamefont
  {Voulgarakis}}\ and\ \bibinfo {author} {\bibfnamefont {J.}~\bibnamefont
  {Chu}},\ }\href@noop {} {\bibfield  {journal} {\bibinfo  {journal} {J. Chem.
  Phys.}\ }\textbf {\bibinfo {volume} {130}},\ \bibinfo {pages} {134111}
  (\bibinfo {year} {2009})}\BibitemShut {NoStop}%
\bibitem [{\citenamefont {Hasimoto}(1959)}]{Hasimoto:1959p589}%
  \BibitemOpen
  \bibfield  {author} {\bibinfo {author} {\bibfnamefont {H.}~\bibnamefont
  {Hasimoto}},\ }\href@noop {} {\bibfield  {journal} {\bibinfo  {journal} {J.
  Fluid Mech.}\ }\textbf {\bibinfo {volume} {5}},\ \bibinfo {pages} {317}
  (\bibinfo {year} {1959})}\BibitemShut {NoStop}%
\bibitem [{\citenamefont {Lindbo}\ and\ \citenamefont
  {Tornberg}(2010)}]{Lindbo:2010ha}%
  \BibitemOpen
  \bibfield  {author} {\bibinfo {author} {\bibfnamefont {D.}~\bibnamefont
  {Lindbo}}\ and\ \bibinfo {author} {\bibfnamefont {A.~K.}\ \bibnamefont
  {Tornberg}},\ }\href@noop {} {\bibfield  {journal} {\bibinfo  {journal} {J.
  Comput. Phys.}\ }\textbf {\bibinfo {volume} {229}},\ \bibinfo {pages} {8994}
  (\bibinfo {year} {2010})}\BibitemShut {NoStop}%
\bibitem [{\citenamefont {Berne}\ and\ \citenamefont
  {Pecora}(1990)}]{bernebook}%
  \BibitemOpen
  \bibfield  {author} {\bibinfo {author} {\bibfnamefont {B.~J.}\ \bibnamefont
  {Berne}}\ and\ \bibinfo {author} {\bibfnamefont {R.}~\bibnamefont {Pecora}},\
  }\href@noop {} {\emph {\bibinfo {title} {Dynamic Light Scattering: with
  applications to Chemistry, Biology and Physics}}}\ (\bibinfo  {publisher}
  {Dover},\ \bibinfo {year} {1990})\BibitemShut {NoStop}%
\bibitem [{\citenamefont {Weeks}\ \emph {et~al.}(1971)\citenamefont {Weeks},
  \citenamefont {Chandler},\ and\ \citenamefont {Andersen}}]{Weeks:1971us}%
  \BibitemOpen
  \bibfield  {author} {\bibinfo {author} {\bibfnamefont {J.~D.}\ \bibnamefont
  {Weeks}}, \bibinfo {author} {\bibfnamefont {D.}~\bibnamefont {Chandler}}, \
  and\ \bibinfo {author} {\bibfnamefont {H.~C.}\ \bibnamefont {Andersen}},\
  }\href@noop {} {\bibfield  {journal} {\bibinfo  {journal} {J. Chem. Phys.}\
  }\textbf {\bibinfo {volume} {54}},\ \bibinfo {pages} {5237} (\bibinfo {year}
  {1971})}\BibitemShut {NoStop}%
\bibitem [{\citenamefont {Jorgensen}\ and\ \citenamefont
  {Madura}(1985)}]{tip4p}%
  \BibitemOpen
  \bibfield  {author} {\bibinfo {author} {\bibfnamefont {W.~L.}\ \bibnamefont
  {Jorgensen}}\ and\ \bibinfo {author} {\bibfnamefont {J.~D.}\ \bibnamefont
  {Madura}},\ }\href@noop {} {\bibfield  {journal} {\bibinfo  {journal} {J.
  Chem. Phys.}\ }\textbf {\bibinfo {volume} {56}},\ \bibinfo {pages} {1381}
  (\bibinfo {year} {1985})}\BibitemShut {NoStop}%
\bibitem [{\citenamefont {Humphrey}\ \emph {et~al.}(1996)\citenamefont
  {Humphrey}, \citenamefont {Dalke},\ and\ \citenamefont {Schulten}}]{vmd96}%
  \BibitemOpen
  \bibfield  {author} {\bibinfo {author} {\bibfnamefont {W.}~\bibnamefont
  {Humphrey}}, \bibinfo {author} {\bibfnamefont {A.}~\bibnamefont {Dalke}}, \
  and\ \bibinfo {author} {\bibfnamefont {K.}~\bibnamefont {Schulten}},\
  }\href@noop {} {\bibfield  {journal} {\bibinfo  {journal} {J. Molec.
  Graphics}\ }\textbf {\bibinfo {volume} {14}},\ \bibinfo {pages} {33}
  (\bibinfo {year} {1996})}\BibitemShut {NoStop}%
\bibitem [{\citenamefont {Berendsen}\ \emph {et~al.}(1984)\citenamefont
  {Berendsen}, \citenamefont {Postma}, \citenamefont {van Gunsteren},
  \citenamefont {DiNola},\ and\ \citenamefont {Haak}}]{berendsen}%
  \BibitemOpen
  \bibfield  {author} {\bibinfo {author} {\bibfnamefont {H.~J.~C.}\
  \bibnamefont {Berendsen}}, \bibinfo {author} {\bibfnamefont {J.~P.~M.}\
  \bibnamefont {Postma}}, \bibinfo {author} {\bibfnamefont {W.~F.}\
  \bibnamefont {van Gunsteren}}, \bibinfo {author} {\bibfnamefont
  {A.}~\bibnamefont {DiNola}}, \ and\ \bibinfo {author} {\bibfnamefont {J.~R.}\
  \bibnamefont {Haak}},\ }\href@noop {} {\bibfield  {journal} {\bibinfo
  {journal} {J. Chem. Phys.}\ }\textbf {\bibinfo {volume} {81}},\ \bibinfo
  {pages} {3684} (\bibinfo {year} {1984})}\BibitemShut {NoStop}%
\bibitem [{\citenamefont {Bussi}\ \emph {et~al.}(2007)\citenamefont {Bussi},
  \citenamefont {Donadio},\ and\ \citenamefont {Parrinello}}]{Bussi:2007p114}%
  \BibitemOpen
  \bibfield  {author} {\bibinfo {author} {\bibfnamefont {G.}~\bibnamefont
  {Bussi}}, \bibinfo {author} {\bibfnamefont {D.}~\bibnamefont {Donadio}}, \
  and\ \bibinfo {author} {\bibfnamefont {M.}~\bibnamefont {Parrinello}},\
  }\href@noop {} {\bibfield  {journal} {\bibinfo  {journal} {J. Chem. Phys.}\
  }\textbf {\bibinfo {volume} {126}},\ \bibinfo {pages} {014101} (\bibinfo
  {year} {2007})}\BibitemShut {NoStop}%
\bibitem [{\citenamefont {Hess}\ \emph {et~al.}(2008)\citenamefont {Hess},
  \citenamefont {Kutzner},\ and\ \citenamefont {van~der Spoel}}]{gromacs4}%
  \BibitemOpen
  \bibfield  {author} {\bibinfo {author} {\bibfnamefont {B.}~\bibnamefont
  {Hess}}, \bibinfo {author} {\bibfnamefont {C.}~\bibnamefont {Kutzner}}, \
  and\ \bibinfo {author} {\bibfnamefont {D.}~\bibnamefont {van~der Spoel}},\
  }\href@noop {} {\bibfield  {journal} {\bibinfo  {journal} {J. Chem. Theor.
  Comput.}\ }\textbf {\bibinfo {volume} {4}},\ \bibinfo {pages} {435} (\bibinfo
  {year} {2008})}\BibitemShut {NoStop}%
\bibitem [{\citenamefont {Weiss}\ \emph {et~al.}(2008)\citenamefont {Weiss},
  \citenamefont {Raschke},\ and\ \citenamefont {Levitt}}]{weiss08}%
  \BibitemOpen
  \bibfield  {author} {\bibinfo {author} {\bibfnamefont {D.~R.}\ \bibnamefont
  {Weiss}}, \bibinfo {author} {\bibfnamefont {T.~M.}\ \bibnamefont {Raschke}},
  \ and\ \bibinfo {author} {\bibfnamefont {M.}~\bibnamefont {Levitt}},\
  }\href@noop {} {\bibfield  {journal} {\bibinfo  {journal} {J. Phys. Chem. B}\
  }\textbf {\bibinfo {volume} {112}},\ \bibinfo {pages} {2981} (\bibinfo {year}
  {2008})}\BibitemShut {NoStop}%
\bibitem [{\citenamefont {Yeh}\ and\ \citenamefont
  {Hummer}(2004)}]{Yeh:2004p287}%
  \BibitemOpen
  \bibfield  {author} {\bibinfo {author} {\bibfnamefont {I.~C.}\ \bibnamefont
  {Yeh}}\ and\ \bibinfo {author} {\bibfnamefont {G.}~\bibnamefont {Hummer}},\
  }\href@noop {} {\bibfield  {journal} {\bibinfo  {journal} {J. Phys. Chem. B}\
  }\textbf {\bibinfo {volume} {108}},\ \bibinfo {pages} {15873} (\bibinfo
  {year} {2004})}\BibitemShut {NoStop}%
\bibitem [{\citenamefont {Gonzalez}\ and\ \citenamefont
  {Abascale}(2010)}]{viscoref}%
  \BibitemOpen
  \bibfield  {author} {\bibinfo {author} {\bibfnamefont {M.~A.}\ \bibnamefont
  {Gonzalez}}\ and\ \bibinfo {author} {\bibfnamefont {J.~F.}\ \bibnamefont
  {Abascale}},\ }\href@noop {} {\bibfield  {journal} {\bibinfo  {journal} {J.
  Chem. Phys.}\ }\textbf {\bibinfo {volume} {132}},\ \bibinfo {pages} {096101}
  (\bibinfo {year} {2010})}\BibitemShut {NoStop}%
\bibitem [{\citenamefont {Schmidt}\ and\ \citenamefont
  {Skinner}(2003)}]{schmidt03}%
  \BibitemOpen
  \bibfield  {author} {\bibinfo {author} {\bibfnamefont {J.~R.}\ \bibnamefont
  {Schmidt}}\ and\ \bibinfo {author} {\bibfnamefont {J.~L.}\ \bibnamefont
  {Skinner}},\ }\href@noop {} {\bibfield  {journal} {\bibinfo  {journal} {J.
  Chem. Phys.}\ }\textbf {\bibinfo {volume} {119}},\ \bibinfo {pages} {8062}
  (\bibinfo {year} {2003})}\BibitemShut {NoStop}%
\bibitem [{\citenamefont {Schmidt}\ and\ \citenamefont
  {Skinner}(2004)}]{schmidt04}%
  \BibitemOpen
  \bibfield  {author} {\bibinfo {author} {\bibfnamefont {J.~R.}\ \bibnamefont
  {Schmidt}}\ and\ \bibinfo {author} {\bibfnamefont {J.~L.}\ \bibnamefont
  {Skinner}},\ }\href@noop {} {\bibfield  {journal} {\bibinfo  {journal} {J.
  Phys. Chem. B}\ }\textbf {\bibinfo {volume} {108}},\ \bibinfo {pages} {6767}
  (\bibinfo {year} {2004})}\BibitemShut {NoStop}%
\bibitem [{\citenamefont {Li}(2009)}]{li09}%
  \BibitemOpen
  \bibfield  {author} {\bibinfo {author} {\bibfnamefont {Z.}~\bibnamefont
  {Li}},\ }\href@noop {} {\bibfield  {journal} {\bibinfo  {journal} {Phys. Rev.
  E}\ }\textbf {\bibinfo {volume} {80}},\ \bibinfo {pages} {061204} (\bibinfo
  {year} {2009})}\BibitemShut {NoStop}%
\bibitem [{\citenamefont {Zangi}(2011)}]{Zangi:2011p581}%
  \BibitemOpen
  \bibfield  {author} {\bibinfo {author} {\bibfnamefont {R.}~\bibnamefont
  {Zangi}},\ }\href@noop {} {\bibfield  {journal} {\bibinfo  {journal} {J.
  Phys. Chem. B}\ }\textbf {\bibinfo {volume} {115}},\ \bibinfo {pages} {2303}
  (\bibinfo {year} {2011})}\BibitemShut {NoStop}%
\bibitem [{\citenamefont {Morrone}\ \emph {et~al.}(2010)\citenamefont
  {Morrone}, \citenamefont {Zhou},\ and\ \citenamefont
  {Berne}}]{Morrone:2010p363}%
  \BibitemOpen
  \bibfield  {author} {\bibinfo {author} {\bibfnamefont {J.~A.}\ \bibnamefont
  {Morrone}}, \bibinfo {author} {\bibfnamefont {R.}~\bibnamefont {Zhou}}, \
  and\ \bibinfo {author} {\bibfnamefont {B.~J.}\ \bibnamefont {Berne}},\
  }\href@noop {} {\bibfield  {journal} {\bibinfo  {journal} {J. Chem. Theor.
  Comput.}\ }\textbf {\bibinfo {volume} {6}},\ \bibinfo {pages} {1798}
  (\bibinfo {year} {2010})}\BibitemShut {NoStop}%
\bibitem [{\citenamefont {Sun}\ and\ \citenamefont {Weinstein}(2007)}]{sun07}%
  \BibitemOpen
  \bibfield  {author} {\bibinfo {author} {\bibfnamefont {J.}~\bibnamefont
  {Sun}}\ and\ \bibinfo {author} {\bibfnamefont {H.}~\bibnamefont
  {Weinstein}},\ }\href@noop {} {\bibfield  {journal} {\bibinfo  {journal} {J.
  Chem. Phys.}\ }\textbf {\bibinfo {volume} {127}},\ \bibinfo {pages} {155105}
  (\bibinfo {year} {2007})}\BibitemShut {NoStop}%
\bibitem [{\citenamefont {Frembgen-Kesner}\ and\ \citenamefont
  {Elcock}(2010)}]{elcock10}%
  \BibitemOpen
  \bibfield  {author} {\bibinfo {author} {\bibfnamefont {T.}~\bibnamefont
  {Frembgen-Kesner}}\ and\ \bibinfo {author} {\bibfnamefont {A.~H.}\
  \bibnamefont {Elcock}},\ }\href@noop {} {\bibfield  {journal} {\bibinfo
  {journal} {Biophys. J.}\ }\textbf {\bibinfo {volume} {99}},\ \bibinfo {pages}
  {L75} (\bibinfo {year} {2010})}\BibitemShut {NoStop}%
\bibitem [{\citenamefont {Zwanzig}(1990)}]{Zwanzig:1990uo}%
  \BibitemOpen
  \bibfield  {author} {\bibinfo {author} {\bibfnamefont {R.}~\bibnamefont
  {Zwanzig}},\ }\href@noop {} {\bibfield  {journal} {\bibinfo  {journal} {Acc.
  Chem. Res.}\ }\textbf {\bibinfo {volume} {23}},\ \bibinfo {pages} {148}
  (\bibinfo {year} {1990})}\BibitemShut {NoStop}%
\end{thebibliography}

\end{document}